\documentclass[pre,aps,reprint,superscriptaddress]{revtex4-2}
\usepackage{graphicx}
\usepackage{dcolumn}
\usepackage{bm}
\usepackage{color}
\usepackage{xcolor}
\usepackage{natbib}
\usepackage[utf8]{inputenc}
\usepackage{float}
\usepackage{amsmath}
\usepackage{amsfonts}
\usepackage{textgreek} 
\usepackage{enumerate}
\usepackage[colorlinks=true,linkcolor=blue,citecolor=red]{hyperref}%

\usepackage{multirow}

\newcommand{\std}{\textrm{std}}
\newcommand{\var}{\textrm{var}}
\newcommand{\cov}{\textrm{cov}}
\newcommand{\diag}{\textrm{diag}}
\newcommand{\trace}{\textrm{Tr}}

\def\N{{\mathcal N}}

\newcommand{\clr}{\color{black}}

\begin{document}

\author{Yann Lanoisel\'ee}
\email{Corresponding author: ylanoiselee@bcamath.org }
\affiliation{
BCAM -- Basque Center for Applied Mathematics,
Alameda de Mazarredo 14, 48009 Bilbao, Basque Country -- Spain}

\author{Denis S. Grebenkov}
\email{Contact author: denis.grebenkov@polytechnique.edu}
\affiliation{ Laboratoire de Physique de la Mati\`ere Condens\'ee (UMR 7643),
CNRS -- Ecole Polytechnique, Institut Polytechnique de Paris, 91120 Palaiseau, France
}

\author{Gianni Pagnini}
\email{Contact author: gpagnini@bcamath.org }
\affiliation{
BCAM -- Basque Center for Applied Mathematics,
Alameda de Mazarredo 14, 48009 Bilbao, Basque Country -- Spain}
\affiliation{
Ikerbasque -- Basque Foundation for Science, Plaza Euskadi 5, 48009 Bilbao, Basque Country -- Spain}

\begin{abstract}
We introduce the concept of 
Randomly Modulated Gaussian Processes as a unifying framework for {\clr elaborating}, analyzing and classifying anomalous diffusion models in {\clr annealed} heterogeneous media. This formulation incorporates correlations in the displacements together with correlated fluctuations of their amplitudes. Most known models of anomalous diffusion (including continuous-time random walk, fractional Brownian motion{\clr , and L\'evy flights}) and random diffusivity can be described and further generalized within this framework. Moreover, the unified view identifies the main statistical properties to be probed experimentally for a reliable classification of diffusive dynamics. The proposed matrix formulation facilitates the computation of the first four moments and allows for a systematic statistical characterization of the considered processes. The necessary and sufficient conditions are provided for the emergence of anomalous diffusion. General expressions for the non-Gaussian parameter, the ergodicity breaking parameter and the covariance of squared increments are derived. {\clr An expression for the characteristic function and the codifference (i.e.,  a generalized measure of correlations) are obtained and used to study the special cases of L\'evy flights and Laplace motion with correlated displacements. Potential} applications of this framework for systematic analysis and biophysical interpretations of experimental single-particle trajectories are discussed.
\end{abstract}
 
\date{\today}
\title{
A unifying approach to diffusive transport in {\clr annealed} heterogeneous media
}
\maketitle
\section{Introduction}
Diffusion in heterogeneous media is a generic feature of mesoscopic physical systems leading to deviations from classical Brownian motion. Spatial and temporal disorder in the local environment couples to thermal fluctuations, giving rise to anomalous scaling laws and non-Gaussian displacement statistics. Such behavior has been observed across a wide range of systems, including glassy and supercooled materials~\cite{POGANY1976,Stariolo2006}, systems close to the jamming transition~\cite{Chaudhuri2007}, cold-atom experiments~\cite{Sagi2012}, complex fluids such as worm-like micellar solutions~\cite{Jeon2013}, active suspensions~\cite{Leptos2009}, and colloidal particles in structured environments~\cite{Chakraborty2020}. Closely related phenomena also arise in biological systems, including search and transport inside living cells~\cite{Bronstein2009,Witzel2019,Sabri2020,Wang2009} and on their membranes~\cite{Weigel2011,Sungkaworn2017-db,Grimes2023-ns}. Moreover, anomalous and non-Gaussian diffusion can emerge even in the absence of active driving, solely due to environmental heterogeneity. Understanding diffusion under these conditions is essential for describing transport-controlled processes such as reaction kinetics, equilibration, and molecular signaling, with implications ranging from soft and condensed matter to biological physics~\cite{Ben-Avraham,Metzler,Lindenberg,Grebenkov}.

Two major aspects of diffusion in heterogeneous media, which may occur independently, are the anomalous scaling of the mean-squared displacement (MSD) $\langle X^2(t) \rangle \propto t^\alpha$, and non-Gaussian displacement statistics. The scaling exponent $\alpha$ distinguishes subdiffusion ($0 < \alpha < 1$), normal diffusion ($\alpha = 1$), and superdiffusion ($1 < \alpha < 2$).
Subdiffusion can originate from long stalling periods between particle displacements, as exemplified by 
the continuous-time random walk (CTRW)~\cite{METZLER2000,Kutner2017}. When the stalling periods are
power-law distributed (CTRW-Pow) anomalous diffusion emerges in close relation to time-fractional diffusion equation
\cite{wyss-jmp-1986,hilfer_etal-pre-1995,mainardi-csf-1996}, {\clr Anomalous diffusion also occurs when the diffusion coefficient deterministically decays as a power-law as seen in the scaled Brownian motion (sBm)~\cite{Jeon2014}.
Finally,} anomalous diffusion can also originate
from the viscoelastic properties of the medium, such as the cytoplasm
of a living cell, where long-range correlations of particle displacements
are captured by fractional Brownian motion
(fBm)~\cite{Mandelbrot1968,Benelli_2021} and the generalized Langevin equation (GLE)~\cite{Zwanzig1973,Porra1996,Bertseva2012,Goychuk2012}. {\clr Finally, anomalous diffusion with infinite second moment occurs in L\'evy flights~\cite{Fogedby1994} when the jump size is power-law distributed.}
Non-Gaussianity is another key aspect of diffusion in heterogeneous media which indicates variability of at least one intensive quantity (e.g., diffusivity) describing the ensemble. This is the basic idea of superstatistics \cite{Beck2003,Beck2005}, which encompasses heterogeneity described as a continuous stochastic process in diffusing diffusivity (DD)
models~\cite{Chubynsky2014,doi:10.1021/acs.jpcb.6b01527,PhysRevE.98.052138,PhysRevX.7.021002,Lanoiselee2018_FPT,sposini_etal-njp-2018,w8gv-3fxt,Sposini2024}, a piecewise-constant function in switching diffusivity (SD) models~\cite{Jensen2005,Fieremans2010,Grebenkov2019,Gueneau2025,Miyaguchi2016,Miyaguchi2019,Balcerek_2023,Pacheco2024} {\clr including free/trapped dynamics~\cite{Doerries2022,Chaudhuri2007,Chakraborty2020}}, or random yet trajectory-wise constant diffusivity (Rand-Const)~\cite{schneider-1990,schneider-1992,mura_etal-jpa-2008}.
Non-Gaussianity appears as a generic property of diffusion in heterogeneous media as illustrated for CTRW with exponential waiting times (CTRW-Exp)~\cite{Barkai2020,Burov2026}. Over the years a large variety of models have been introduced to describe these statistical features (see~\cite{Akimoto_2026} for further discussion).

These models employ different mathematical tools such as Fokker-Planck equations, renewal theory, subordination, fractional calculus, or Gaussian statistics that are difficult to combine into a unified description of general correlated random walks in heterogeneous media. 

In this paper, we fill this gap and introduce a theoretical framework for diffusive motion in heterogeneous media that captures both anomalous scaling and non-Gaussian fluctuations within a unified description.
The proposed framework naturally incorporates correlations in the
displacements, non-Gaussianity due to medium heterogeneity, anomalous power-law scaling of the MSD, deterministic or random changes in statistical properties, and other features. 


{\clr The paper is organized as follows. In Sec.~\ref{sec:RMGP} we introduce the model of Randomly Modulated Gaussian Processes (RMGPs). When the position moments exist, we compute exactly the MSD (Sec.~\ref{sec:Cov_MSD}), the non-Gaussian parameter (Sec.~\ref{sec:NG_param} and Appendix~\ref{appendix:Extended_non_Gaussian}), the Ergodicity Breaking (EB) parameter (Sec.~\ref{sec:Ergodicity_breaking_param}), and the covariance of squared increments (Sec.~\ref{sec:cov_squared_inc}).
This analysis uncovers three fundamental quantities that encode all information about the dynamics up to the fourth moment.
We also derive an expression for the characteristic function of RMGPs (see Sec.~\ref{sec:characteristic_function}) and their codifference (Sec.~\ref{sec:Codifference}) which allow one to consider cases when the moments do not exist and to study distributional properties of large-jump processes such as L\'evy flights or the Laplace motion with correlated displacements. We then discuss some limitations and interpretations and conclude the manuscript with the summary of main results, open problems and perspectives (Sec.~\ref{sec:Conclusion}).
}
\section{Randomly Modulated Gaussian Processes}\label{sec:RMGP}
{\clr 
To get a physical motivation, we consider the motion of particles in a visco-elastic medium subject to annealed disorder understood as adiabatic fluctuations of an intensive quantity (temperature, viscosity, hydrodynamic radius, etc.) in the system. Fast moving components (e.g. water) induce Gaussian displacements (thermal noise). In addition, the disorder in the medium may lead to fluctuations of the amplitude of each displacement that we call ``random modulations''. The amplitudes may be correlated as timescales as long as the time it takes for a particle to fully sample the disorder. The visco-elastic nature of the medium induces correlations of the displacements. When moments of random modulations are finite, we show that the knowledge of their mean and covariance is sufficient to deduce the first four moments of the position.}

Bearing this physical picture in mind,
we introduce the framework of RMGPs, 
in which the amplitudes of Gaussian increments are randomly rescaled by a stochastic process,
that {\clr resembles the notion of stochastic volatility in mathematical finance and }extends the existing terminology
of a randomly scaled Gaussian process~\cite{sliusarenko_etal-jpa-2019}.
This framework produces a sequence of successive random positions at discrete time steps and thus facilitates the analysis and interpretation of experimentally acquired trajectories, as well as computer-based simulations. {\clr We emphasize that the underlying process can also be 
defined in continuous time (see Sec. \ref{sec:Conclusion}), but the presentation is more intuitive with sampling at discrete times.}
In mathematical terms, if $\delta$ denotes a time step (fixed by the acquisition technique or chosen in simulations), we represent a random trajectory $X(t)$ of the diffusing particle by an $N$-dimensional vector $X$, whose components $X_i$ are the successive positions $X(i\delta)$ at times $i\delta$, with $i=1,2,\ldots,N$. A matrix representation of Gaussian processes is used to incorporate correlations of displacements through a covariance matrix. In addition, the Gaussian jumps are rescaled to mimic heterogeneity as may be described by CTRW or random diffusivity models~\cite{mura_etal-jpa-2008,Chubynsky2014}. For clarity of presentation, we focus on one-dimensional settings but its extension to higher dimensions is immediate when the process coordinates are independent.

The construction of a random trajectory involves two sets of $N$ random variables: (i) the 
standard independent identically distributed Gaussian variables $\{\xi_1,\cdots,\xi_N\}$ to represent thermal noises, and (ii) the positive random variables $J_i$ to mimic the effect of heterogeneity. 
The random variables $\{J_1,\cdots,J_N\}$ are independent of $\{\xi_1,\cdots,\xi_N\}$ such that $\langle \sqrt{J_i} \xi_j\rangle=\langle \sqrt{J_i} \rangle\langle\xi_j\rangle=0$ for any $i$ and $j$ {\clr (where $\langle\cdot\rangle$ denotes the ensemble average)}; in turn, eventual correlations between the random modulations $J_i$ are allowed.

At this stage, the modulated Gaussian variables $\sqrt{J_i}\xi_i$ are first-order uncorrelated, $\langle (\sqrt{J_i}\xi_i) (\sqrt{J_j}\xi_j)\rangle = \langle J_i\rangle\delta_{ij}$ but second-order correlated: $\langle (\sqrt{J_i} \xi_i)^2 (\sqrt{J_j} \xi_j)^2\rangle = \langle J_i J_j\rangle$ for $i\neq j$ and $\langle (\sqrt{J_i} \xi_i)^4\rangle = 3\langle J_i ^2\rangle$ reflecting the kurtosis of the Gaussian distribution $\langle \xi_i^4\rangle=3$. 

Next, for a given positive-definite matrix $C$, {\clr with its square root verifying $C=\sqrt{C}\sqrt{C}^\top$}, we define the $n$-th increment as
$Y_n=\sum_{k=1}^n (\sqrt{C})_{nk}\sqrt{J_k}\xi_k$. In the classical case without modulations (i.e., $J_i=1$), one has $\langle Y_j Y_k\rangle = C_{j,k}$, i.e., $C$ {\clr is} the covariance matrix of Gaussian increments. In the presence of modulations, this is not true anymore; {\clr nevertheless} $C$ is used to induce first-order correlations of increments (in the following we still keep calling it `covariance matrix'). 

\begin{figure}[!t]
\includegraphics[width=\columnwidth]{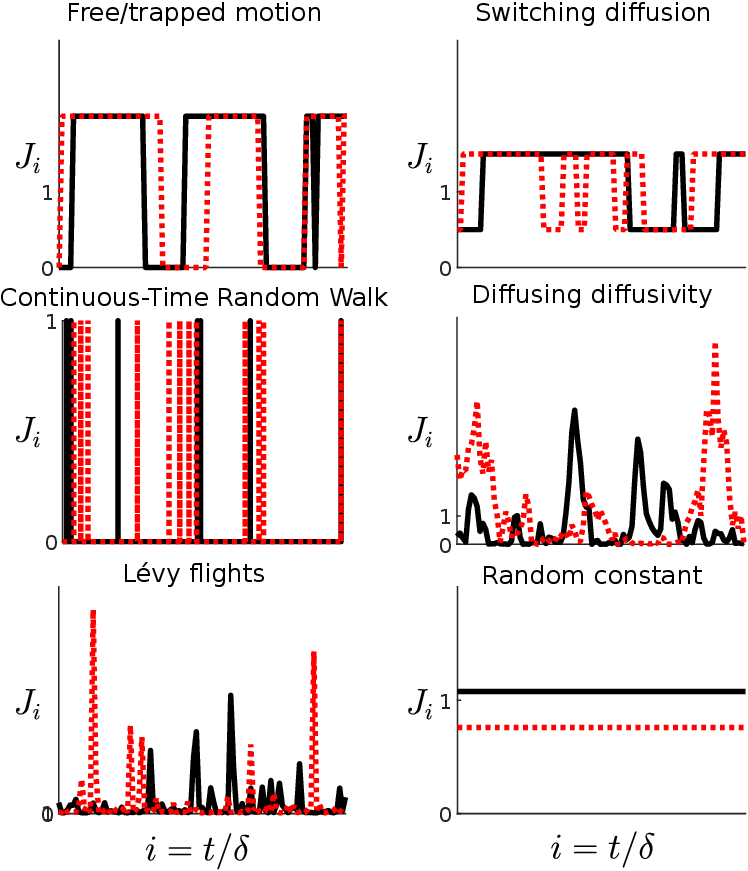}
\caption{
{\clr
Two independent realizations (solid-black and dashed-red) of random modulations for some popular processes.
}
}\label{fig:example_modulation}
\end{figure}

The positions are finally obtained by summing the increments: $X_n=\sum_{i=1}^nY_i$ with $n\in[1,2,\ldots,N]$. First-order correlations control the behavior of the MSD, while second-order correlations determine the fourth moment and thus control non-Gaussianity, as detailed below.
The model can be represented in a matrix form as
\begin{equation}\label{eq:LJE_def}
 X = L \,\sqrt{C} \, \sqrt{J}\, \xi ,
\end{equation}
where $L$ is a lower triangular matrix with entries $1$ performing integration, $C$ is the covariance matrix {\clr(see Appendix~\ref{appendix:how_to_get_C_and_J} for its construction)}, $J$ is the diagonal matrix containing random modulations, and $\xi$ is the vector of standard IID Gaussian random variables. 
As the positions in $X$ should be in length units, we set $\xi$ and $J$ to be unitless, while $C$ has units of a squared length. 
The matrix representation (\ref{eq:LJE_def}), which is the key point of the paper, is very general and includes as special cases numerous models studied in the literature.
In particular, we retrieve:
\begin{enumerate}[(i)]
    \item  Brownian motion: $C \propto I$ and $J= I$, where $I$ is the identity matrix.
\item Gaussian processes: $J=I$, including fBm and GLE.
\item CTRW: $C \propto I$.  In fact, $J_{i}$ can be understood as the number of jumps (or renewals) between two position recordings $(i-1)\delta$ and $i\delta$, whose asymptotic PDF has been studied~\cite{Schulz2013,Barkai2003}. {\clr In Appendix~\ref{appendix:CTRW}, we present the properties of $J$ for CTRW. }
\item Random Constant (Rand-Const): setting $J = \mu I$, one can reproduce a statistical ensemble of random walks, in which each
trajectory is multiplied by a random scale $\mu$ to describe anomalous diffusion and everlasting non-Gaussianity. 
The two most studied examples are the grey Brownian motion (gBm)~\cite{schneider-1990,schneider-1992} and generalized grey Brownian motion
(ggBm)~\cite{mura_etal-jpa-2008,molina_etal-pre-2016},
that allow one to retrieve a wide range of 
fractional governing equations 
\cite{runfola_etal-pd-2024} 
including time-fractional diffusion. 
\item Switching diffusivity (SD) corresponds to
 modulations induced by a piecewise-constant stochastic process along the diagonal of $J$. 
The random waiting times between switching events can be exponentially distributed~\cite{Jensen2005,Fieremans2010,Grebenkov2019,Gueneau2025} (SD-Exp) or power-law distributed 
(SD-Pow)~\cite{Miyaguchi2016,Miyaguchi2019,Balcerek_2023,Pacheco2024}. A special case is the Annealed Transit-Time Model (ATTM)~\cite{Massignan2014,Manzo2015,Akimoto_2016} where the random state duration $\tau$ is correlated with the random diffusivity $D$ in the form $\tau\propto D^{-\alpha}$, with some exponent $\alpha$. 
\item Diffusing Diffusivity (DD): modulations modeled by a positive stochastic process (e.g., many works were devoted to diffusing diffusivity
with exponentially decaying correlations (DD-Exp) when diffusivity
fluctuates around its mean~\cite{Chubynsky2014,PhysRevX.7.021002,Lanoiselee2018_FPT}). {\clr In Appendix~\ref{appendix:square_gaussian_process}, we consider diffusing diffusivity modeled as the square of an arbitrary Gaussian process, and compute its expectation and covariance. Random modulations defined as the square of a Gaussian process have a gamma-like distribution which typically induces exponential tails to the position PDF~\cite{Kozachenko1999}. Additionally, it allows modeling a general class of covariance structures for $J$ when the expectation is constant (see Appendix~\ref{appendix:how_to_get_C_and_J}). 
}
\item {\clr L\'evy Flights: when random modulations are power-law distributed (see Sec.~\ref{sec:Levy_flight}).}
\end{enumerate}
{\clr
Moreover, all processes with $C\propto I$ (e.g. Rand-const, SD, DD, CTRW, and L\'evy flights) admits an alternative interpretation as Brownian motion with a random operational time $S(n)$~
\cite{Sposini2024a,Sposini2024b,Lanoiselee2018_NGmodel,PhysRevX.7.021002,Magdziarz2008}, which is given in our framework as the sum of random modulations: $S(n)=\sum_{i=1}^nJ_{i}$ with $n\in[1,2,\ldots,N]$.}
In Figure~\ref{fig:example_modulation}, we show examples of the modulations for several types of motion. This illustrates the generality of the random modulation representation of annealed disorder.

\begin{figure}[!t]
\includegraphics[width=\columnwidth]{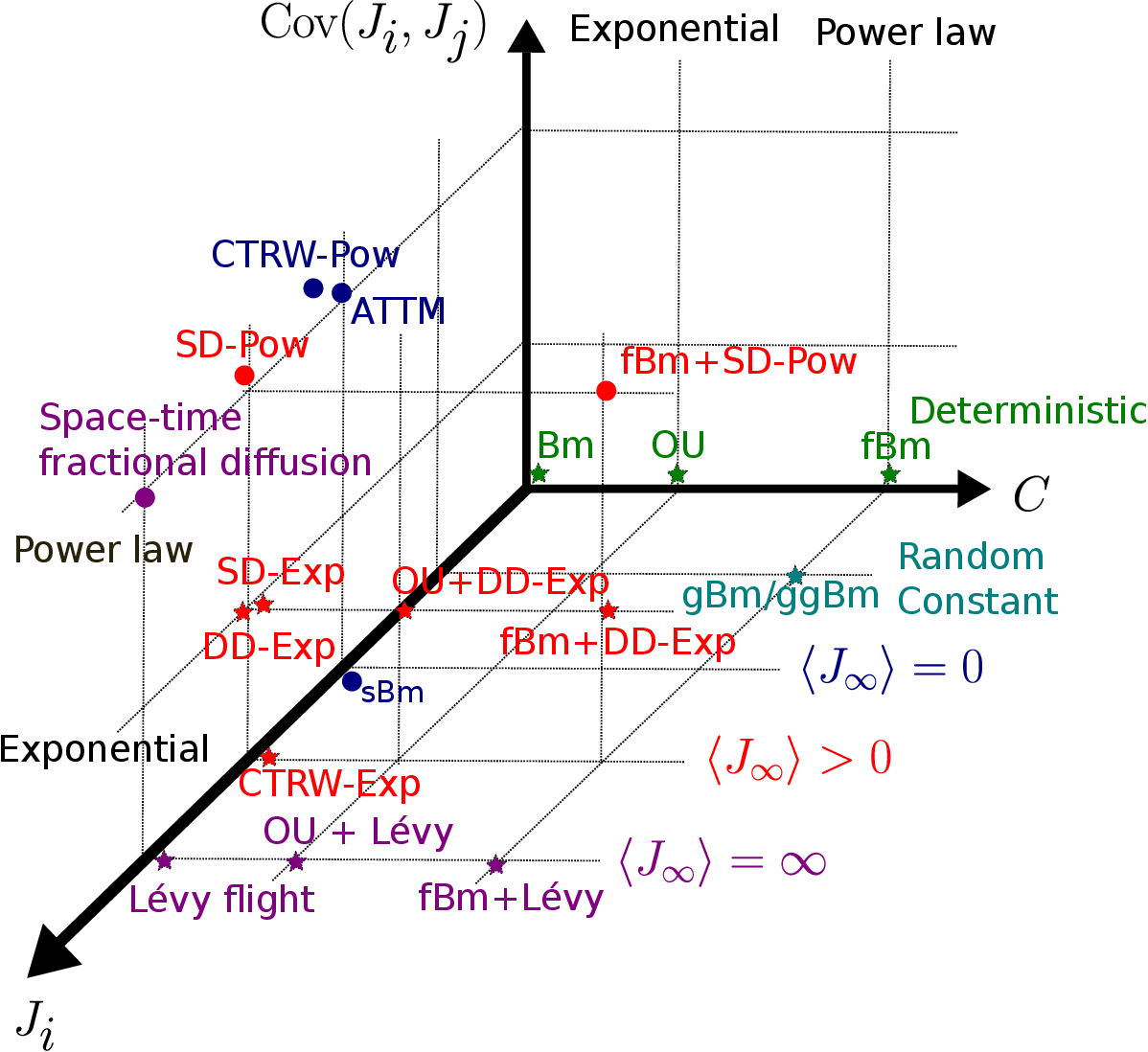}
\caption{
Three-dimensional diagram of annealed diffusion models in heterogeneous media. The direction $C$ corresponds to first-order correlations of displacements going from uncorrelated (at the origin) to exponentially and power-law correlated. The direction $J_i$ corresponds to the type of modulations, whether they are deterministic, random but constant, {\clr fluctuating with freezing, fluctuating with positive long-time limit, or fluctuating with infinite mean}.
Finally, the $\cov(J_i,J_j)$ direction corresponds to correlations of modulations, ranging from uncorrelated (near origin) to exponentially and power-law correlated. Stars denote processes with trajectory-wise stationary increments (meaning that both $C$ and $\cov(J_i,J_j)$ are stationary) and disks denote those with non-stationary increments. The abbreviated models are described in the text.}
\label{fig:model_space}
\end{figure}

While the above list is far from exhaustive, the diversity of the presented models reflects the complexity of experimentally observed diffusive behaviors and highlights the need for a unified framework for their classification and comparison.
The RMGPs defined through Eq.~(\ref{eq:LJE_def}) provides such a unifying framework, organizing the realm of diffusion models within a three-dimensional parameter space (Fig.~\ref{fig:model_space}). 
\begin{enumerate}
\item
The first axis represents first-order correlations, encoded in the matrix $C$, ranging from (i) Markovian dynamics (no correlation), (ii) exponentially correlated processes (e.g., Ornstein-Uhlenbeck (OU) process), to (iii) power-law correlated processes (e.g., fBm).
\item
The second axis characterizes the nature of modulations $J_i$, which may be
(i) deterministic (e.g., Bm, OU, fBm), 
(ii) a random constant (e.g., gBm, ggBm), 
(iii) a freezing process with
$\langle J_\infty \rangle = 0$
(e.g., CTRW-Pow, ATTM), 
(iv) a stochastic process with a positive asymptotic mean
$\langle J_\infty \rangle > 0$
(e.g., CTRW-Exp, DD-Exp, SD-Exp, SD-Pow), 
(v) {\clr a stochastic process with infinite mean $\langle J_\infty \rangle = \infty$ (e.g., L\'evy flights)}. 
\item 
The third axis corresponds to correlations of random modulations, which may be (i) absent (e.g., CTRW-Exp), (ii) exponentially decaying (e.g., DD-Exp, SD-Exp), and (iii) power-law decaying (e.g., CTRW, ATTM, SD-Pow).
\end{enumerate}

The practical value of our unifying representation~(\ref{eq:LJE_def}) goes far beyond the classification of known models. Various combinations of the physical effects can be simultaneously realized by changing the properties of $C$ and $J$. For example, DD modulation, which was originally proposed for Brownian motion, can be applied to 
under-damped Langevin dynamics~\cite{vitali_etal-jrsi-2018,sliusarenko_etal-jpa-2019}, to
the Ornstein-Uhlenbeck process~\cite{Lanoiselee2023} (that we call 
OU+DD-Exp in Fig.~\ref{fig:model_space}), and to fractional Brownian motion~\cite{Wang2020} (fBm+DD-Exp).
{\clr Similarly, generalized versions of L\'evy flights with correlated displacements can be constructed such as L\'evy fractional stable motion (fBm+L\'evy) ~\cite{Burnecki2010}, and L\'evy-driven Ornstein-Uhlenbeck process (OU+L\'evy)~\cite{Maller2009}. In this way, one can produce a variety of new models with well-controlled statistical properties.} As experimental data do not necessarily follow a particular model, the diagram in Fig.~\ref{fig:model_space} invites rethinking the statistical analysis of the data beyond the restricted framework of specific models. While a lot effort was put into attributing experimental data to a specific model, we encourage to classify experimental datasets according to the statistical properties of $C$ and $J$, as illustrated in Fig. \ref{fig:model_space}. {\clr Development of reliable inference tools for such classifications presents an important direction for future research.}

The advantage of our discrete-time framework is
threefold: first, the matrix representation allows one to incorporate the building blocks of the dynamics in a general way and to get an exact derivation of its first four moments; second, the discrete integration is performed through multiplication by the $L$ matrix, which is a simple operation that can easily be inverted (with $L^{-1}$ corresponding to differentiation). {\clr Finally, the characteristic function of RMGPs admits a formal expression that can be exactly solved in some relevant cases.}

In heterogeneous media, the distribution of the position of a particle may display exponential tails. While models explaining exponential tails often suppose that the diffusivity varies in time, some universality results have
been derived based on CTRW with exponential waiting times
(CTRW-Exp)~\cite{Barkai2020,Wang2020b}. Our formulation reveals that CTRW-Exp is a process with random yet uncorrelated modulations (see Appendix~\ref{appendix:CTRW}). 
Upon time rescaling, RMGPs with rapidly decaying correlations of their modulation (e.g., exponential or faster) can be matched to CTRW-Exp for which the tail behavior is known~\cite{Burov2026}.

{\clr To illustrate the efficiency and versatility of our framework, we examine the respective roles of the three ingredients that are required to characterize the dynamics up to the fourth moment: the covariance
matrix $C$, the expectation $\langle J_i\rangle$, and the
covariance of random modulations $\cov(J_i,J_j)$.
We will obtain the exact representations for the main statistics and discuss their links to the known models.
}
\section{Statistical properties for finite moments}
{\clr In this section, we compute a number of important descriptors used in the analysis of diffusive transport. Those descriptors are (i) the Mean-Squared Displacement and the covariance matrix to investigate space exploration, (ii) the non-Gaussian parameter to characterize eventual thermalization of the system, (iii) the ergodicity breaking parameter to quantify reliability of measurement based on the Time-Averaged Mean-Squared Displacement, and (iv) the covariance of square increments which may give access to the covariance matrix of random modulations. }

\subsection{Covariance and MSD}
\label{sec:Cov_MSD}
The covariance of RMGPs can be obtained by evaluating two averages: over thermal noises $\xi$ and over random
modulations $J$ (average over disorder), from which
\begin{equation}
 \langle X \, X^\top\rangle = L \, \sqrt{C} \, \langle J\rangle \, \sqrt{C}^\top \, L^\top.
\end{equation}
The covariance matrix of the process depends thus on $C$ and on the expectation of modulations.
While the MSD vector can be directly deduced from the diagonal of the
covariance matrix, it is informative to represent the MSD at time
$t=N\delta$ as
\begin{equation} \label{eq:MSD_general}
 \langle X^2(t)\rangle=\trace \left( \langle J \rangle U \right),
\end{equation}
where $U=\sqrt{C}^\top S \sqrt{C}$ with $S = L + L^\top- I$
being a $N\times N$ matrix with all entries equal to $1$. {\clr The diagonal elements $U_{ii}$ represent the time-derivative of the MSD at step $i$ obtained from the unmodulated process (with the same $C$ but with constant $J_i=1$). Expression (\ref{eq:MSD_general})} shows that in our framework anomalous scaling can originate either from a power-law scaling of the covariance matrix (e.g., fBm), or from non-stationary random modulations whose expectation exhibits a power-law decay (e.g., scaled Brownian motion (sBm) or CTRW-Pow) {\clr depending on whether it is $U_{ii}$ or $\langle J_i\rangle$ that scales as $(i\delta)^{\alpha-1}$ or both, where $\alpha$ is the anomalous scaling exponent}. 
For instance, if random modulations are initiated from an equilibrium state such that
$\langle J_i\rangle=\langle J \rangle$ for all $i\in[1,\ldots,N]$, the contribution of
random modulations can be factored out. Applying the random modulation to
fBm-type covariance of increments with a generalized diffusion coefficient $D_\alpha$ and the Hurst exponent $H=\alpha/2$ leads to the MSD
\begin{equation} 
 \langle X^2(t)\rangle = 2\langle J \rangle D_\alpha t^{\alpha}. 
\end{equation}
We conclude that the fluctuations of random modulations around their mean play
no role in the behavior of the MSD; however they are relevant
for non-Gaussianity (see below).
Figure~\ref{fig:statistics}.A shows the MSD for seven representative models of anomalous diffusion: fBm, CTRW-Exp, DD-Exp, SD-Exp, sBm, CTRW-Pow, and Rand-Const. 
Perfect agreement between theory and simulation can be observed.
{\clr We distinguish two cases: (i) the MSD has subdiffusive scaling when either $U_{ii}$ (fBm) or $\langle J_i\rangle$ (CTRW-Pow and sBm) scales as $(i\delta)^{\alpha-1}$, and (ii) the MSD is linear in time in all other cases because both $U_{ii}$ and $\langle J_i\rangle$ are constants.}

\begin{figure}[!htbp]
\includegraphics[width=\columnwidth]{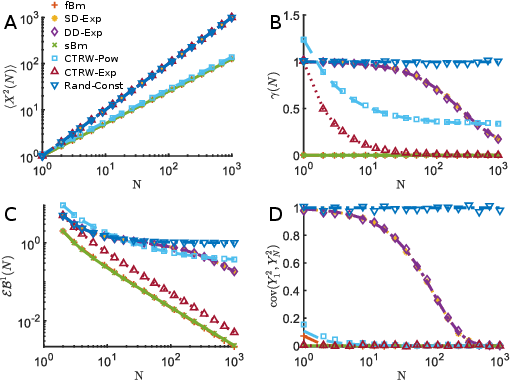}
\caption{(A) Mean-squared displacement as a function of time. (B) Non-Gaussian
parameter as a function of time. (C) Ergodicity breaking parameter as a function of the number $N$ of steps in a trajectory. (D) Autocovariance of squared increments as a function of time. Lines show theoretical predictions, whereas symbols present simulations that are obtained for each model by generating $M=10^6$ trajectories. {\clr Seven models are compared: fBm (with $H=0.35$ and $J_i=1$), DD-Exp (with $\langle D\rangle=1$ with $\tau=10$), SD-Exp (with two states $D_1=0$, $D_2=2$ and switching probabilities $p_1 = p_2 = 0.005$), sBm (with $\alpha=0.7$ and $\sigma^2=1$), CTRW-Pow (with $\alpha=0.7$ and $\sigma^2=1$), CTRW-Exp (with $\lambda=1$ and $\sigma^2=1$), and random constant modulations with $J$ exponentially distributed with scale $1$}. See Appendix~\ref{appendix:simulation details} for more information on simulation procedures and model parameters.
}
\label{fig:statistics}
\end{figure}


\subsection{Non-Gaussian parameter}
\label{sec:NG_param}

Next, we consider the non-Gaussian parameter
\begin{equation}
 \gamma(t)=\frac{1}{3}
 \frac{\langle X^4(t)\rangle}{\langle X(t)^2\rangle^2}-1,
\end{equation}
which quantifies deviations from a Gaussian distribution. 
According to Eq. (\ref{eq:LJE_def}), for arbitrary $C$ and $\cov(J_i,J_j)$ we get the general representation,
\begin{equation}\label{eq:NG_param}
\gamma(t)=
 \frac{
 \sum\nolimits_{i,j=1}^NU_{ii} \,\cov(J_i,J_j)\,U_{jj}
 }{
 \bigl(\sum\nolimits_{i=1}^N \langle J_{i} \rangle U_{ii} \bigr)^2
 },
\end{equation}
where $U$ is defined just after Eq.~(\ref{eq:MSD_general}). 
This representation allows us to discuss the conditions under which non-Gaussianity
emerges, and those leading to long-time thermalization and ultimate convergence
to Gaussianity. Within the framework of RMGPs, the necessary and sufficient condition
for the PDF to be non-Gaussian is to have random modulations; in turn,
if the modulations are deterministic, the PDF is necessarily Gaussian
(e.g., sBm). The correlations of the increments enter in the
non-Gaussian parameter through the matrix $U$.
Figure~\ref{fig:statistics}.B 
presents the non-Gaussian parameter for fBm, CTRW-Exp, DD-Exp, SD-Exp, sBm, CTRW-Pow, and Rand-Const, and illustrates perfect agreement between theory and simulation. 
{\clr This figure illustrates the diversity of non-Gaussianity profiles. When $J\propto I$, the process is Gaussian (fBm). When the random modulations are independent, we get $1/t$ convergence to Gaussian (CTRW-Exp). For exponentially correlated models (DD-Exp and SD-Exp), the PDF remains non-Gaussian for times shorter than the correlation timescale and then converges to a Gaussian as $1/t$. In turn, models with non-stationary power-law decaying correlations of modulations (CTRW-Pow) never converge to a Gaussian, although one can observe non-constant $\gamma(t)$ at short times until the asymptotic regime is reached. Finally, processes with random but trajectory-wise constant modulations (Rand-Const) exhibit constant $\gamma(t)$, self averaging is not possible in this case. }
For first-order uncorrelated displacements ($C\propto I$), one has $U_{ij}=\sigma^2$
{\clr for all $i$ and $j$,} where $\sigma$ is a lengthscale. If in addition $\langle J_i\rangle=\langle J\rangle$ and $\cov(J_i,J_j)=f(|i-j|)$ with a decreasing function $f$, Gaussian statistics is necessarily recovered as $t\to\infty$ due to thermalization (see Appendix~\ref{appendix:Extended_non_Gaussian}). For exponentially correlated modulations, $f(|i-j|) \propto \exp(-|i-j|\delta/t_c)$, the second and fourth moments exactly coincide with those of DD-Exp and SD-Exp models~\cite{Fieremans2010,doi:10.1021/acs.jpcb.6b01527,PhysRevX.7.021002,sposini_etal-njp-2018,Lanoiselee2018_NGmodel,Grebenkov2019,Gueneau2025} {\clr (see Fig.~\ref{fig:statistics}.B)}. In turn, for second-order uncorrelated displacements ($\cov(J_i,J_j)\propto \delta_{i,j}$), one gets $\gamma(t)\propto (\delta/t) \, \var(J)/\langle J\rangle^2$, yielding the characteristic $1/t$ convergence to a Gaussian distribution~\cite{Lanoiselee2019}, with CTRW-Exp as an example (see Appendix~\ref{appendix:CTRW}); exponentially correlated DD-Exp models, as well as non-Gaussian diffusion near surfaces~\cite{Alexandre2023}, display the same asymptotic behavior for $t\gg\tau$.
In Appendix~\ref{appendix:Extended_non_Gaussian}, we discuss in details the relationship to previously published literature, and the conditions under which the PDF may remain non-Gaussian at any time.



\subsection{Ergodicity breaking parameter}
\label{sec:Ergodicity_breaking_param}

When considering diffusion in heterogeneous media, it is important to
check whether a single trajectory is representative of the ensemble.
In other words, one should assess the ergodic behavior of the system. Within RMGPs, weak ergodicity breaking can be addressed systematically. For this purpose, one can compare the MSD to its time-averaged counterpart, the time-averaged mean-squared displacement (TAMSD) at lag-time $\tau$: $M_N(\tau)=\frac{1}{N-\tau}\sum\limits_{n=1}^{N-\tau}(X_{n+\tau}-X_n)^2$. To have ergodicity, two ingredients are required. First, the time average and the ensemble average should be equal in the long-time limit. Second, the time average should converge with probability one to the ensemble average, thus meaning the PDF of the time-average should converge to a Dirac delta function $\delta(M_N(\tau)-\langle X^2(\tau)\rangle)$ as $N\to\infty$. The TAMSD can be
expressed as the quadratic form
\begin{equation}
 M_N(\tau)=X_N^\top R_{\tau} X_N,
\end{equation}
where $R_{\tau}$ is a matrix representing the
lag-differences~\cite{Grebenkov2011b,Lanoiselee2025}. The expectation can be computed using the trace trick and the law of total expectation
\begin{equation}\label{eq:TAMSD_expectation}
 \langle M_N(\tau)\rangle =\trace(\langle J\rangle V_\tau),
\end{equation}
where $V_{\tau}=\sqrt{C}^\top L^\top R_{\tau} L\sqrt{C}$ and we recall that $L$ is a lower triangular matrix with all non-zero elements being ones. Looking at
the first condition, the expectation of random modulations should
converge to a constant as $N\to\infty$. Another way to formulate this condition
is to say that the initial condition should not influence the long-time
behavior. For instance, the dependence of the Ornstein-Uhlenbeck
process on its initial condition decays
exponentially fast. While the expectation of $\langle
J_{i}\rangle$ is not time invariant, the
process is ergodic in the weak sense because the asymptotic behavior of the MSD is independent of the initial condition.

To test the second condition, the ratio between the variance and the squared expectation of the time-average has been studied under the name of ``ergodicity-breaking parameter'', which can be computed within the framework of RMGPs.
From the trace trick and the law of total variance, one can show that the variance of TAMSD is
\begin{equation}
 \var (M_N(\tau)) =2\trace(\langle J V_\tau J V_\tau \rangle)+\var(\trace(J V_\tau)),
\end{equation}
where 
\begin{equation}
\begin{split}
 \var(\trace(J V_\tau))&=\left\langle\sum\limits_{i,j} J_iV_{ii}J_jV_{jj}\right\rangle-\left\langle\sum\limits_{i} J_iV_{ii}\right\rangle^2
 \\
 &=\sum\limits_{i,j} V_{ii}V_{jj}\cov( J_iJ_j),
\end{split}
\end{equation}
 from which one can express the ergodicity-breaking parameter for a lag-time $\tau$:
\begin{equation}
 \mathcal{EB}^\tau=\frac{ 
 \sum\limits_{i,j=1}^N (2 V^\tau_{ij} V^\tau_{ji}+V^\tau_{ii}V^\tau_{jj})\,\cov( J_i,J_j) +
 2\langle J_i \rangle\langle J_j \rangle V^\tau_{ij} V^\tau_{ji}
 }{
 \left(\sum\limits_{i,j=1}^NV^\tau_{ii}\langle J_i\rangle\right)^2},
\end{equation}
where $V_\tau$ is defined just below Eq.~(\ref{eq:TAMSD_expectation}). Figure~2.C shows perfect agreement between theory and simulation for seven representative models: fBm, CTRW-Exp, DD-Exp, SD-Exp, sBm, CTRW-Pow, and Rand-Const. Usually,
$\mathcal{EB}^\tau$ is plotted for a fixed lag-time as a function of $N$. For uncorrelated displacements at
$\tau=1$ we have $V_{ij}=\delta_{ij}(1-\delta_{i1})\,\sigma^2/(N-1)$, such that $\mathcal{EB}^1$ can be expressed as 
\begin{equation}
 \mathcal{EB}^1=
\frac{\sum_{i=2}^N 2(\var(J_i)+\langle J_i\rangle^2)+\sum_{i,j=2}^N\cov(J_i,J_j)}{\left(\sum_{i=2}^N\langle J_i\rangle\right)^2} . 
\end{equation}
When additionally the random modulations are stationary (i.e., $\langle J_i\rangle=\langle J\rangle$ and $\var(J_i)=\var(J)$) we get
\begin{equation}
 \mathcal{EB}^1=
 \frac{2}{N-1}\left(1+\frac{ \var(J)}{\langle J\rangle^2}\right)
 +
 \frac{ 
 \sum\limits_{\substack{i,j=2 \\ i\neq j}}\cov( J_i,J_j)
 }{
 (N-1)^2\langle J\rangle^2},
\end{equation}
where the covariance of random modulations has been splitted into diagonal and non-diagonal parts. 
The first term corresponds to the
$1/N$ decay, known for Brownian motion. The term $\var(J)/\langle J\rangle^2$ (the diagonal part of the covariance matrix) which accounts for uncorrelated random
modulations, does not change the scaling as a function of $N$ but modifies its amplitude. The third term accounts for correlations of random
modulations, which may dominate at short times. For example, when the covariance decays exponentially, i.e.,
$\cov(J_i,J_j)=\var(J)e^{-|i-j| \delta/t_c}$ and $N\delta<t_c$,
$\mathcal{EB}^1$ can be almost constant. Therefore, TAMSD measurements on short trajectories displaying correlated random modulations are highly unreliable.


\subsection{Covariance of squared increments}
\label{sec:cov_squared_inc}

As random modulations play a crucial role in this framework, it is important to be able to access their covariance. In this context, we consider the covariance of squared increments. 
Denoting $K = \sqrt{C}$, we get 
\begin{eqnarray}\nonumber
\cov( Y_i^2 , Y_j^2 ) 
&=&
\sum\limits_{k,p=1}^N \Big[\cov(J_{k},J_{p}) (K_{ik}K_{jp})^2 \left(1+\frac{2K_{jk}K_{ip}}{K_{ik}K_{jp}}\right)\\
&&
+2\langle J_k\rangle\langle J_p \rangle K_{ik}K_{ip}K_{jk}K_{jp}\Big],
\end{eqnarray}
for $i\neq j$, and 
 $\var(Y_i^2)=3\sum\nolimits_{k,p=1}^N (K_{ik}K_{ip})^2\cov(J_{k},J_{p})$, 
where $K = \sqrt{C}$.
Figure~\ref{fig:statistics}.D presents this covariance for fBm, CTRW-Exp, DD-Exp, SD-Exp, sBm, CTRW-Pow, and Rand-Const models. When the
displacements are uncorrelated (i.e., $C\propto I$), this expression is reduced to $\cov( Y_i^2 , Y_j^2 ) =\sigma^4\cov(J_{i},J_{j})$
for $i\ne j$, and $\var(Y_i^2)=3\sigma^4\,\var(J_{i})$ for $i=j$, so that it gives a direct measure of the covariance of random modulations. In turn, when the modulations are deterministic ($\cov(J_i,J_j)=0$) one has the amazing property $\cov( Y_i^2 , Y_j^2 )=2\,\cov( Y_i , Y_j )^2$, independently of $C$ and $\langle J_i\rangle$. This property can be tested on experimental data to uncover the nature of modulations.
{\clr 
\section{Characteristic function}
\label{sec:characteristic_function}
While so far our emphasis was on the moments, here we investigate the characteristic function of RMGPs that allows one to understand the distribution of the position $X_N$. For a particular realization of the process, one can interpret a single realization of RMGP as a Gaussian process with a random covariance matrix $\Sigma_J$ such that $X=\sqrt{\Sigma_J}\xi$. The randomness of the covariance matrix is due to the random modulations $J_i$:
\begin{equation}
 \Sigma_J = G\,J\,G^\top,
\end{equation}
where $G=L\sqrt{C}$.
The characteristic function of $X_N$ conditioned on the modulations $J_i$ 
is
\begin{equation}
 \left \langle \exp\left(ikX_{N}\right)|J_1,\ldots,J_N\right\rangle =\exp\left(-\frac{k^2}{2}\sum_{i=1}G_{Ni}J_{i}G_{iN}^\top \right).
\end{equation}
To obtain the characteristic function one has to integrate over all $J_i$ with their joint probability $P(J_1,\ldots,J_N)$:
\begin{equation}\label{eq:char_fun_def}
\begin{split} 
& \left \langle \exp\left(ikX_{N}\right)\right\rangle=
\int_0^\infty dJ_1\ldots\int_0^\infty dJ_N
\\&
\exp\left(-\frac{k^2}{2}\sum_{i=1}G_{Ni}J_{i}G_{iN}^\top\right)P(J_1,\ldots,J_N).
\end{split}
\end{equation}
The characteristic function of the vector $X$ can be obtained in a similar fashion. 
This highlights that the characteristic function requires full information on the first-order correlation contained in $C$ (or $G$) but also the whole distribution of random modulations.
\subsection{Correlated Laplace motion}
For uncorrelated modulations with exponential distribution $P(J_i=x)=\frac{1}{\lambda}e^{-x/\lambda}$ but first-order correlated displacements (through $C$), we get
\begin{equation}
 \left \langle \exp\left(ikX_{N}\right)\right\rangle
 =\prod_{i=1}^N\frac{1}{1+\frac{1}{2\lambda}G_{Ni}G_{iN}^\top k^2}.
 \end{equation}
 At step $N=1$ the distribution is Laplace; the distribution at larger times is obtained by multiple convolutions of Laplace distributions. 
One can study the tail behavior by looking at large displacements (i.e., $k\to 0$) for which only the terms in $k^2$ survive in the denominator:
\begin{equation}
 \left \langle \exp\left(ikX_{N}\right)\right\rangle
 \approx\frac{1}{1+\frac{1}{2\lambda}\sum_{i=1}^NG_{Ni}G_{iN}^\top k^2}
 =\frac{1}{1+\frac{1}{2\lambda}\Sigma_{NN} k^2},
 \end{equation}
 with $\Sigma=LCL^\top$ being the covariance of the unmodulated process (with $J_i=1$). This corresponds to the characteristic function of the Laplace distribution with a MSD given by $\langle X^2(t)\rangle=\frac{1}{2\lambda}\Sigma_{NN}$.
 
 When $C\propto I$ (therefore $G\propto L$) it reduces to
\begin{equation}
 \left \langle \exp\left(ikX_{N}\right)\right\rangle
 =\frac{1}{\left(1+\frac{1}{2\lambda}\sigma^2 k^2\right)^N},
 \end{equation}
 which is Laplace distributed at short times (small $N$) and converges to a Gaussian as $N\to\infty$:
 \begin{equation}
 \left \langle \exp\left(ikX_{N}\right)\right\rangle
 \simeq\exp\left(-\frac{\sigma^2}{2\lambda}N k^2\right),
 \end{equation}
where $\frac{\sigma}{\sqrt{\lambda}}$ is the typical lengthscale of a displacement.
\subsection{Correlated symmetric L\'evy flights}
\label{sec:Levy_flight}
An important special class of heterogeneous diffusion is the L\'evy flight with heavy-tailed displacement distribution. This model can be reproduced by using $J_i$ with power-law distribution, for which we obtain the characteristic function.
More specifically, we consider independent $J_i$ with a one-sided L\'evy distribution with parameter $\alpha$ with Laplace transform
\begin{equation} \label{eq:laplace_transform}
\mathcal{L}\left\lbrace P(J_i=x)\right\rbrace=\int_0^\infty e^{-xs}P(x)dx=e^{-s^{\alpha}}.
\end{equation}
Computing Eq.~(\ref{eq:char_fun_def}) given Eq.~(\ref{eq:laplace_transform}) we get the characteristic function
\begin{equation}\label{eq:Char_Fun_Levy_flight}
 \left \langle \exp\left(ikX_{N}\right)\right\rangle=\exp\left(-\sum_{i=1}^N\left(\frac{1}{2}G_{Ni}G_{iN}^\top\right)^\alpha |k|^{2\alpha}\right).
 \end{equation}
In this way, we obtain a generalized version of the symmetric L\'evy flight with correlated displacements (through $C$). In particular, we retrieve two important examples: Fractional L\'evy stable motion~\cite{Burnecki2010} (by taking $C$ from fractional Gaussian noise), and L\'evy-driven Ornstein-Uhlenbeck process~\cite{Maller2009} (by taking $C$ from the increments of an OU process).

\subsection{Codifference}
 \label{sec:Codifference} Codifference is a tool that extends the concept of correlation to processes with infinite moments~\cite{WYLOMANSKA2015}. It is defined as
 \begin{equation} \mathrm{CD}(X_{n_1},X_{n_2})=\ln\langle e^{iX_{n_2}}\rangle+\ln\langle e^{-iX_{n_1}}\rangle-\ln\langle e^{i(X_{n_2}-X_{n_1})}\rangle.
 \end{equation}
For Gaussian processes this reduces to $-\cov(X_{n_1},X_{n_2})$.
As $\var(X_{n_2}-X_{n_1})=\var (X_{n_2})+\var (X_{n_1}) -2\cov (X_{n_2},X_{n_1})$, the characteristic function of the difference, conditioned on $J_i$, is 
\begin{equation}
\begin{split}
 &\langle  e^{ik(X_{n_2}-X_{n_1})}|J_1,J_2,\ldots,J_N\rangle=
 \exp\biggl(-\frac{k^2}{2}  \\ 
 & \times  \sum_{i=1}^N (G_{n_1,i}J_iG_{i,n_1}^\top
+G_{n_2,i}J_iG_{i,n_2}^\top -2 G_{n_2,i}J_iG_{i,n_1}^\top)\biggr).
 \end{split}
\end{equation}

Factoring the terms $J_i$ and integrating over each of them gives
\begin{equation}
 \begin{split}
 &\langle e^{ik(X_{n_2}-X_{n_1})}\rangle= 
 \exp\biggl(- \frac{|k|^{2\alpha}}{2^\alpha} \\
 & \times \sum_{i=1}^N\bigl( G_{n_1,i}G_{i,n_1}^\top 
 +G_{n_2,i}G_{i,n_2}^\top -2G_{n_2,i}G_{i,n_1}^\top\bigr)^\alpha\biggr).
 \end{split}
\end{equation}
 For correlated L\'evy flights with the characteristic function from Eq.~(\ref{eq:Char_Fun_Levy_flight}) we get
\begin{equation}\label{eq:CD_Levy_flight}
 \begin{split}
&\mathrm{CD}(X_{n_1},X_{n_2})=
\\&
 - \frac{1}{2^\alpha} \sum_{i=1}^{N}\biggl\{\left(G_{n_2i}G_{in_2}^\top\right)^\alpha
 -
 \left(G_{n_1i}G_{in_1}^\top\right)^\alpha
 \\&
 +
\left( G_{n_1,i}G_{i,n_1}^\top+G_{n_2,i}G_{i,n_2}^\top-2G_{n_2,i}G_{i,n_1}^\top \right)^\alpha \biggr\},
 \end{split}
 \end{equation}
which is consistent with the computations in~\cite{WYLOMANSKA2015}.
The special case $\alpha=1$ reduces, as expected, to 
\begin{equation}\label{eq:CD_Levy_flight}
 \mathrm{CD}(X_{n_1},X_{n_2})=
 - \cov(X_{n_2},X_{n_1}).
 \end{equation}
In Fig.~\ref{fig:codifference}, we present the codifference for the L\'evy-driven OU process with $\alpha=0.75,0.875,1$ (red, blue, and purple lines, respectively) and fixed $\theta=1/30$ as well as for fractional L\'evy stable motion with $H=0.3,0.5, 0.7$ (red, blue, and purple lines, respectively) and fixed $\alpha=0.8$. For the L\'evy-driven OU process, the codifference decreases exponential fast, with a decay slowing down with smaller $\alpha$. For fractional L\'evy stable motion, we distinguish three cases: (i) uncorrelated motion for $H=0.5$  (i.e., $\mathrm{CD}=0$), (ii) anticorrelated at long times for $H=0.35$ (i.e., $\mathrm{CD}>0$) and, (iii) positively correlated at long times (i.e., $\mathrm{CD}<0$) for $H=0.7$.
\begin{figure}[!htbp]
\includegraphics[width=\columnwidth]{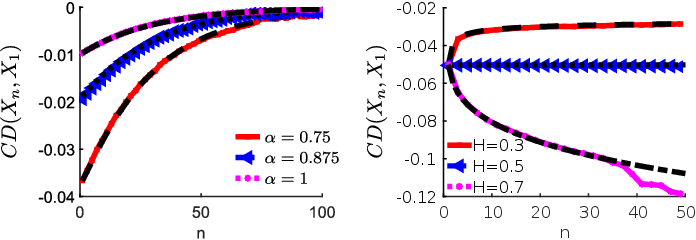}
\caption{Codifference for (left) L\'evy-driven OU process for $\alpha=0.75,0.875,1$ and fixed $\theta=1/30$, and (right) fractional L\'evy stable motion with $H=0.3,0.5, 0.7$ and fixed $\alpha=0.8$.
Black lines are theoretical predictions.}
\label{fig:codifference}
\end{figure}
}

\section{Conclusion}
\label{sec:Conclusion}
To conclude, we
proposed randomly modulated Gaussian processes as a general model of
diffusion in {\clr annealed} heterogeneous media driven by temperature fluctuations.
This approach unifies a broad class of anomalous and/or non-Gaussian diffusion. The related models are characterized up to their fourth moment by the knowledge of
three statistical quantities: the covariance matrix $C$ determining first-order correlations, the
expectation of random modulations, and their covariance. Our representation clarifies the relationships among different models and explains why they
share some statistical properties and not others. {\clr We also obtain the characteristic function of correlated L\'evy flights and their codifference.}
The RMGP model in Eq.~(\ref{eq:LJE_def}) can be viewed as a discrete
version of the integral representation of the continuous in time stochastic process
\begin{equation}
 X(t)=\int_0^t M(t,t')\sqrt{J(t')} \, dW_{t'} \,,
\end{equation}
where $M(t,t')$ is a memory kernel and $\sqrt{J(t)}$ is a time-dependent random process. {\clr We stress that the present description is limited to annealed disorder where the spatial structure of disorder is neglected. As such, effects related to spatial structure~\cite{Pacheco2023,Bouchaud1990} are not included in the phenomenology.}

This work opens the possibility to treat more general diffusive processes and may have a number of extensions as, for example, ggBm-like approaches for the whole spectrum of solutions provided by space-time fractional diffusion \cite{pagnini_etal-fcaa-2016}. 
Large-deviation approaches~\cite{Barkai2020,Wang2020b,Pacheco2024b} may provide {\clr the description of rare events and} distributional properties for RMGPs. Moreover, {\clr our framework suggests a potential} link between diffusion in heterogeneous media and random matrix theory, opening the possibility of reconstructing $C$ and $\langle J\rangle$ directly from empirical data~\cite{Laloux2000,Lamrani2025}. {\clr Extension to randomized covariance matrix $C$ might allow to retrieve processes with random Hurst exponent~\cite{Balcerek2025}.}

{\clr 
When the displacements are uncorrelated ($C\propto I$), random modulations cannot be interpreted more precisely than `fluctuations of the diffusion coefficient' as it is impossible to distinguish between temperature and friction fluctuations. In turn, for correlated displacements, the physical interpretation of random modulations can be clarified.} Suppose the random modulations originate from time-dependent diffusivity and the fluctuation-dissipation theorem holds. The diffusion coefficient $D=k_BT/\nu$ is the ratio of two quantities that may vary: the temperature $T$ and the friction $\nu$. When diffusion occurs in a potential $V(x)$, the drift term is rescaled by friction but is independent of $T$ that only affects fluctuations. Therefore, {\clr in the presence of a potential} our formulation corresponds to temperature fluctuations. Future work will address extensions to heterogeneous media where the random modulations originate from friction fluctuations. 

\begin{acknowledgments}
{\clr We thank Vittoria Sposini for insightful discussions.} GP and YL acknowledge the support by the Basque Government through the
BERC 2022--2025 program and by the Ministry of Science and Innovation:
BCAM Severo Ochoa accreditation CEX2021-001142-S / MICIN / AEI /
10.13039/501100011033.
\end{acknowledgments}


\appendix


\section{Additional results for non-Gaussian parameter}
\label{appendix:Extended_non_Gaussian}

\subsection{The non-Gaussian parameter for some special cases}
\label{sec:special_cases_NG_param}
In this section we discuss some special cases of the non-Gaussian (NG) parameter from Eq.~(\ref{eq:NG_param}). When displacements are uncorrelated, the matrix $C$ is diagonal with elements $C_{ii}=\sigma^2$, where $\sigma$ is a lengthscale. The quantity $\sigma^2$ can be interpreted as the product of the diffusion coefficient and the timestep: $\sigma^2=D\delta$. Based on the definition of $U$ just below Eq.~(\ref{eq:MSD_general}), one can realize that all the elements of $U$ take the same value $U_{ij}=\sigma^2$. The NG parameter thus simplifies to
\begin{equation}\label{eq:gamma_no_corr}
 \gamma(t)=
\frac{\sum\limits_{i,j=1}^N\cov(J_i,J_j)}{(\sum\limits_{i}\langle J_i\rangle)^2}.
\end{equation}
When a RMGP has a constant expectation and a stationary covariance of the form
\begin{equation}\label{eq:cov_J_stat}
 \cov(J_i,J_j)=\var(J)f(|i-j|),
\end{equation}
the convergence to a Gaussian distribution as $N\to\infty$ necessarily occurs. 
This is the case for diffusing diffusivity at equilibrium for which both the expectation and variance of modulations are constant: $\langle J_i\rangle=\langle J\rangle$, $\var(J_i)=\var(J)$. Under this condition, Eq. (\ref{eq:gamma_no_corr}) simplifies to
\begin{equation}\label{eq:gamma_corr_J}
 \gamma(t)=
 \frac{1}{t^2\langle J\rangle^2}\sum\limits_{i,j=1}^N\cov(J_i,J_j).
\end{equation}

Note that if $J$ is exponentially distributed, one has $\frac{\var(J)}{\langle J\rangle^2}=1$, so that $\gamma(N=1)=1$. This value of the NG parameter corresponds to the Laplace distribution (i.e., $p(x) = e^{-|x|/R }/(2R)$, with some $R > 0$). This distribution has been observed experimentally~\cite{Wang2009} and motivated the introduction of diffusing diffusivity models.

In general when the covariance can be written as in Eq.~(\ref{eq:cov_J_stat}), one gets
\begin{equation}
\begin{split}
\gamma(t)=&\frac{\var(J)}{t^2\langle J\rangle^2}\sum_{i,j=1}^N f(|i-j|)\\
=&\frac{\var(J)}{t\langle J\rangle^2}\left(f(0)+2\sum_{k=1}^{N-1}\left(1-\frac{k}{N}\right)f(k) \right).
\end{split}
\end{equation}
This quantity necessarily vanishes for the whole class of stationary random modulations, because $f(|i-j|)$ is a decreasing function, so thermalization occurs. However the convergence rate to a Gaussian distribution depends on the correlation function $f(|i-j|)$.

For instance, if $f(|i-j|)=\exp(-|i-j|\delta/t_c)$ with some correlation time $t_c$, we get by replacing sums with integrals and denoting $t=N\delta$:
\begin{equation}
 \gamma(t)\approx 2t_c \frac{\var(J)}{t\langle J\rangle^2}\left( 1-\frac{t_c}{t}(1-e^{- t/t_c})\right).
\end{equation}
The approximation becomes exact in the limit $\delta\to 0$ while keeping $t=N\delta$ constant.
This result corresponds exactly to diffusing diffusivity models \cite{PhysRevX.7.021002,Lanoiselee2018_NGmodel}. 
This result is also consistent with a model where the diffusion coefficient alternates between two values $D_1$ and $D_2$ through a Markov chain as in~\cite{Gueneau2025,Fieremans2010}, or even an arbitrary number of states \cite{Grebenkov2019}. In fact, the models DD-Exp and SD-Exp (when started at equilibrium) are indistinguishable up to the first four moments~\cite{Lanoiselee2018_NGmodel}.

A further simplification arises when random modulations are uncorrelated, in which case 
\begin{equation}
 \gamma(t)=\frac{1}{t}\frac{\var(J)}{\langle J\rangle^2}.
\end{equation}
We retrieve the $1/t$ convergence to a Gaussian distribution, as observed for random uncorrelated diffusivity models~\cite{Lanoiselee2019}. This is also found for the long-time behavior of diffusing diffusivity models~\cite{PhysRevX.7.021002,sposini_etal-njp-2018,Lanoiselee2018_NGmodel} and non-Gaussian diffusion near surfaces~\cite{Alexandre2023}.

\subsection{Conditions for everlasting non-Gaussianity}
However, it is also possible that the non-Gaussian parameter does not vanish. For instance, it happens if the PDF is obtained by superimposing trajectories whose modulations are constant over time but randomly distributed across trajectories. This is notably the case of the ggBm that consists in an ensemble of fBms with random but constant diffusion coefficients. For each trajectory $m\in[1,\ldots,M]$, the random diffusion coefficient can be defined as $D^{(m)}=D_0J^{(m)}$, where $D_0$ is a constant and $J^{(m)}$ is a modulation that is trajectory-wise constant but randomly distributed among trajectories. In this case, we note that $\cov(J_i,J_j)=\var(J)$, and the non-Gaussian parameter is constant,
\begin{equation}
 \gamma(t)=\frac{\var(J)}{\langle J\rangle^2},
\end{equation}
independently of the covariance matrix of displacements.

Everlasting non-Gaussianity may also hold when displacements are uncorrelated and random modulations are non-stationary,
with $\sum\limits_{i,j}\cov(J_i,J_j)\propto(\sum\limits_{i}\langle J_i\rangle)^2$. 
For uncorrelated displacements, RMGPs can be viewed as time changed Brownian motions $X(N)=B(S(N))$ with a random operational time $S(N)=\sum_{n=1}^N J_n$. The expectation of the operational time is $\langle S(N)\rangle=\sum_{n=1}^N \langle J_n\rangle$ and its variance is $\var( S(N))=\sum_{i,j=1}^N \cov(J_i,J_j)$.
The NG parameter can therefore be recast as
\begin{equation}
 \gamma(N)=\frac{\var(S(N))}{\langle S(N)\rangle^2}.
\end{equation}
To get everlasting non-Gaussianity, it is sufficient that $\langle S(N)\rangle\propto\std(S(N))$. A noteworthy example is the CTRW-Pow for which, as shown in Sec.~\ref{sec:expec_var_tot_num_jump}, the expectation of the number of jumps has the same scaling $t^\alpha$ as its standard deviation.

Finally, everlasting non-Gaussianity also occurs when the position has a stationary distribution and the correlation of its displacements decays faster than that of random modulations as seen for diffusion in a harmonic potential with fluctuating temperature~\cite{Lanoiselee2023}. The temperature fluctuation case studied in~\cite{Lanoiselee2023} can be recovered by RMGPs assuming equilibrium initial condition and zero mean for the position by choosing $C$ that corresponds to the covariance of increments of the Ornstein-Uhlenbeck process. Given the covariance of the position of the OU process $\langle X(i\delta)X(j\delta)\rangle=\bar{D}\tau (e^{-|i-j|\delta/\tau_x}-e^{-|i+j|\delta/\tau_x})$, where $\bar{D}$ is the diffusion coefficient and $\tau_x$ is the position correlation time (see Appendix~\ref{appendix:how_to_get_C_and_J}), the modulations correspond to the Cox-Ingersoll-Ross process as described in Sec.~\ref{sec:sim_DD-Exp}. 
In this case, the second moment $\langle X^2(t)\rangle$ converges to $2\bar{D}\tau_x$ at long times. In turn, the limit of the fourth moment is $\langle X^4(t\to\infty)\rangle=3\bar{D}^2\tau_x^2\left(1+\frac{1}{\nu(1+\mu)}\right)$, where $1/\nu$ measures the strength of diffusivity fluctuations and the initial shape of the distribution and $\theta$ is half the ratio between the position correlation time and the diffusivity correlation time.
Therefore 
\begin{equation}
\gamma(t\to\infty)=\frac{1}{\nu(1+\theta)},
\end{equation}
which either vanishes when diffusivity is constant ($\nu\to\infty$) or when position correlation time is much larger than the diffusivity one $(\theta\to\infty)$ (see~\cite{Lanoiselee2023} for a complete derivation).

{\clr 
\section{Random modulation of a known Gaussian process}
\label{appendix:how_to_get_C_and_J}

From a Gaussian process with zero mean and known covariance matrix $\Sigma$, the matrix $C$ is obtained by double discrete-time differentiation, which can be computed efficiently in matrix form:
\begin{equation}
 C=L^{-1}\,\Sigma\,{L^{-1}}^\top,
\end{equation}
where $L^{-1}$ is a discrete differential operator.
Alternatively it can be obtained by computing the covariance of increments:
\begin{equation}
 C(i,j)=\langle (Y(i\delta)-Y((i-1)\delta))(Y(j\delta)-Y((j-1)\delta))\rangle,
\end{equation}
 where $Y$ is the unmodulated Gaussian process. In the case of fractional Brownian motion, $C$ is the covariance of fractional Gaussian noise, see Eq.~(\ref{eq:C_for_fBm}).

In Appendix~\ref{appendix:square_gaussian_process} we will show how to approximate random modulations as the square of a Gaussian variable. The covariance of modulations is related to the covariance matrix $Q$ of the underlying Gaussian process to be squared, via:
\begin{align}\nonumber
\cov(J_i,J_j)&=4\langle J\rangle Q_{ij}+2Q_{ij}^2.
\end{align}
Solving the quadratic equation for $Q_{ij}$ and selecting the positive root we get
\begin{equation}
Q_{i,j}=\langle J\rangle\left(\sqrt{1+\frac{\cov(J_i,J_j)}{2\langle J\rangle^2}}-1\right).
\end{equation}
As a consequence, for a desired $\cov(J_i,J_j)$ one can construct the corresponding covariance matrix of the Gaussian process to be squared.
}
\section{Modulation statistics from renewal counting process}
\label{appendix:CTRW}
In this section we discuss how to match a CTRW to RMGPs through the characterization of random modulations that occur in the CTRW. 
In a standard CTRW, a particle executes independent random jumps, drawn from a given PDE $p(x)$, each jump taking an independent random waiting time, drawn from a given PDE $\psi(t)$~\cite{METZLER2000}.

{\clr For clarity, we suppose that $p(x)$ is Gaussian; note that a combination of random modulations with Gaussian jumps allows one to deal with more general distributions $p(x)$.} In turn, many choices are possible for $\psi(t)$. We distinguish two scenarios: (i) when all moments of $\psi(t)$ are finite (e.g. for an exponential distribution), the CTRW is known to be reduced to normal diffusion with a non-Gaussian distribution at short times; (ii) when the mean waiting-time is infinite due to a power-law heavy tail of $\psi(t)$ at long times, the CTRW leads to anomalous diffusion and remains non-Gaussian at any time.

{\clr 
For a single realization of CTRW in the interval $(0,t)$, let $\N_t$ be the random number of jumps realized at random times $t_1,...,t_{\N_t}$. One can define
\begin{equation}
 I(t)=\sum_{k=1}^{\N_t} \delta(t-t_k),
\end{equation}
where each Dirac delta function accounts for a jump.
The trajectory of CTRW can be then subdivided in $N$ steps of duration $\delta$. Setting the random modulation at step $i$ to be the number of jumps between $(i-1)\delta$ and $i\delta$:
\begin{equation}
 J_i=\int\limits_{(i-1)\delta}^{i\delta}I(t')dt',
\end{equation}
the CTRW trajectory can be represented as a discrete random walk:
\begin{equation}
 X(t)=\sum_{i=1}^N\sigma\sqrt{J_i}\xi_i,
\end{equation}
with $\sigma$ being the typical lengthscale of displacements. This establishes a clear reframing of the CTRW as a RMGP.
}

We are interested in finding the expectation and covariance of random modulations $J(t)$ that are required to describe the random position up to the fourth moment. It is easier to start from the number $\N(t)$ of jumps performed up to time $t$ whereas the random modulations can then be obtained as 
\begin{equation}\label{eq:def_modulation_from_S}
 J(t)=\N(t)-\N(t-\delta).
\end{equation} 
Note that the concept of operational time $S(t)$ is very close to the number of jumps upon multiplication by the duration $T$ of a single jump $S(t)=\N(t)T$.
While the simulations are performed in discrete time, we derive the statistical properties of random modulations in the continuous-time framework. For conciseness we use the notation $J(t)$ throughout this section keeping in mind that $J(t)$ depends on $\delta$.

\subsection{Expectation and variance of the number of jumps}\label{sec:expec_var_tot_num_jump}

To obtain the expectation and the variance of modulations, we evaluate the first and second moments of the number $\N(t)$
of jumps realized during a time interval $(0,t)$. 
Let $\tilde{\psi}(s)$ be the Laplace transform of the waiting time density $\psi(t)$ defined by 
\begin{equation}
 \tilde{\psi}(s)=\int_0^\infty e^{-st}\psi(t)dt.
\end{equation}

Let $p_n(t)$ be the probability of having $n$ jumps on the interval (0,t): $p_n(t) = {\mathbb P}\{\N(t) = n\}$. Using a standard renewal argument, one can evaluate the Laplace transform of this probability:
\begin{equation}
\tilde{p}_n(s) = [\tilde{\psi}(s)]^n\frac{1-\tilde{\psi}(s)}{s}.
\end{equation}

As a consequence, we get the Laplace transform of the mean number of jumps as
\begin{equation}
{\mathcal L}\{ \langle {\mathcal N}(t)\rangle \} = \sum\limits_{n=1}^\infty n \, \tilde{p}_n(s) = \frac{\tilde{\eta}(s)-1}{s} \,,
\end{equation}
where $\tilde{\eta}(s) = 1/(1-\tilde{\psi}(s))$, from which 
\begin{equation}\label{eq:expec_jump_number}
\langle \N(t)\rangle = -1 + \int\nolimits_0^{t} dt' \eta(t').\end{equation} 
This
expression gives a useful interpretation for the function $\eta(t)$ as a sort of effective rate of realizing a jump at time $t$. Similarly, one gets the second moment as
\begin{equation}
{\mathcal L}\{ \langle {\mathcal N}^2(t)\rangle \} = \sum\limits_{n=1}^\infty n^2 \, \tilde{p}_n(s) 
= \frac{1 - 3\tilde{\eta}(s) + 2[\tilde{\eta}(s)]^2}{s} \,.
\end{equation}
To model a power-law-decaying waiting-time distribution we chose the Mittag-Leffler distribution with the PDF $\psi(t)=-\partial_t E_\alpha(-(t/T)^\alpha)$, where $E_\alpha(z)$ is the Mittag-Leffler function $E_\alpha(z)=\sum_{n=0}^\infty z^n/\Gamma(1+n\alpha )$, while $T$ is a timescale for a single jump. For this choice, one has 
\begin{equation}\label{eq:psi_s_ML}
 \tilde{\psi}(s) = \frac{1}{1 + (sT)^\alpha},
\end{equation} 
and thus 
\begin{equation}\label{eq:eta_s_ML} 
\tilde{\eta}(s) = 1 +
\frac{1}{(sT)^\alpha},
\end{equation}
 so the expectation and the second moment are
\begin{equation}\label{eq:mean_number_jump_powerlaw}
 \langle \N(t)\rangle =\frac{
t_1^\alpha}{T^\alpha \Gamma(\alpha+1)}
\end{equation}
and
\begin{equation}
\langle \N^2(t)\rangle = \frac{t^\alpha}{T^\alpha \Gamma(\alpha+1)} + \frac{2t^{2\alpha}}{T^{2\alpha} \Gamma(2\alpha+1)} \,.
\end{equation}
For $\alpha = 1$, these expressions give $\langle \N(t)\rangle = t/T$ and
$\langle \N^2(t)\rangle = t/T + (t/T)^2$ so that the
variance of $\N(t)$ is $t/T$, as expected for a Poisson
process. In turn, the $(t/T)^{2\alpha}$ term is still present in
the variance for $\alpha < 1$ so that the standard deviation scales as
$(t/T)^\alpha$, in the same way as the mean value, at large $t$.
In other words, the values of $\N(t)$ are widely spread and not peaked around the mean in the anomalous setting.

\subsection{Two-point moment}
To compute the fourth moment of a random walker performing a RMGP, one needs the covariance $\cov(J(t_1),J(t))$ with $t_1<t$. This requires the computation of the joint statistics of two random variables: the number ${\mathcal N}_1(t_1)$ of jumps in the interval $(0,t_1)$, and the number ${\mathcal N}_2(t_2)$ of jumps in the interval $(t_1,t_1+t_2)$.
Let $P_{n_1,n_2}(t_1,t_2)$ be the probability of having $n_1$ jumps in the
interval $(0,t_1)$ and $n_2$ in the interval $(t_1,t_1+t_2)$. The double Laplace transform of this probability,
\begin{equation}
 \tilde{\tilde{P}}_{n_1,n_2}(s_1,s_2)=\int_0^\infty dt_1\int_0^\infty dt_2 e^{-s_1t_1-s_2t_2}P_{n_1,n_2}(t_1,t_2)
\end{equation}
was obtained in~\cite{Lanoiselee2016} (see Eq. (B9)):
\begin{equation}
\begin{split}
\tilde{\tilde{P}}_{n_1,n_2}(s_1,s_2) =& [\tilde{\psi}(s_1)]^{n_1} \frac{\tilde{\psi}(s_2)-\tilde{\psi}(s_1)}{s_1-s_2}
\\ 
&  \times[\tilde{\psi}(s_2)]^{n_2-1}
\frac{1-\tilde{\psi}(s_2)}{s_2}.
\end{split}
\end{equation}
We get
then the double Laplace transform of the two-point moment $Q(t_1,t_2)=\langle \N_1(t_1)\N_2(t_2)\rangle$ under the form
\begin{equation}
\begin{split}
 \tilde{\tilde{Q}}(s_1,s_2) = &\sum\limits_{n_1,n_2=1}^\infty n_1 n_2 \tilde{\tilde{P}}_{n_1,n_2}(s_1,s_2) \\
 =& [\tilde{\eta}(s_1) - 1 ] \, \frac{\tilde{\eta}(s_2) - \tilde{\eta}(s_1)}{s_1-s_2} \frac{1}{s_2} \,,
\end{split}
\end{equation}
where $\tilde{\eta}(s_2) = 1/(1 - \tilde{\psi}(s_2))$. Note that the
double inverse Laplace transform of $\frac{\tilde{\eta}(s_2) -
\tilde{\eta}(s_1)}{s_1-s_2}$ is simply $\eta(t_1+t_2)$. 

For instance, if $\psi(t) = e^{-t/T}/T$, one has $\tilde{\psi}(s)
= 1/(1 + sT)$ and thus $\tilde{\eta}(s) = 1 + 1/(sT)$, so that
$\tilde{\tilde{Q}}(s_1,s_2) = 1/(T^2 s_2^2 s_1^2)$, from which $Q(t_1,t_2)
= t_1 t_2/T^2$. This is expected for an exponential (Poisson) process. 

For the Mittag-Leffler distribution, Eq.~(\ref{eq:eta_s_ML}) implies
\begin{equation}
\tilde{\tilde{Q}}(s_1,s_2) = \frac{1}{s_2 (s_1T)^\alpha} \biggl( \frac{1/(sT)^\alpha - 1/(s_1T)^\alpha}{s_1-s_2} \biggr).
\end{equation}
As the double inverse Laplace transform of the second factor is
$(t_1+t_2)^{\alpha-1}/(T^\alpha \Gamma(\alpha))$, we obtain
$Q(t_1,t_2)$ as the integral over $t_2$ and the fractional integral over
$t_1$:
\begin{widetext}
\begin{align*} 
& Q(t_1,t_2) = \int\limits_0^{t_2} dt'_2 \int\limits_0^{t_1} dt'_1 \frac{(t_1 - t'_1)^{\alpha-1}}{T^\alpha \Gamma(\alpha)} \, 
\frac{(t'_1+t'_2)^{\alpha-1}}{T^\alpha \Gamma(\alpha)}
= \frac{1}{\alpha T^{2\alpha} \Gamma^2(\alpha)}
\int\limits_0^{t_1} dt'_1 (t_1 - t'_1)^{\alpha-1} \bigl[(t'_1+t_2)^{\alpha} - (t'_1)^\alpha\bigr] \\ 
 = & \frac{t_1^{2\alpha}}{T^{2\alpha} \, \alpha \Gamma^2(\alpha)}
\int\limits_0^1 dz (1-z)^{\alpha-1} \bigl[(z+t_2/t_1)^{\alpha} - z^\alpha\bigr] 
=  \frac{t_1^{\alpha} (t_1+t_2)^\alpha}{T^{2\alpha} \, \alpha^2 \Gamma^2(\alpha)} 
~ _2F_1(-\alpha,\alpha; \alpha+1; 1/(1 + t_2/t_1))- \frac{t_1^{2\alpha}}{T^{2\alpha} \Gamma(2\alpha+1)} \\
 = & \frac{t_1^{2\alpha}}{T^{2\alpha} \, \alpha \Gamma^2(\alpha)} \biggl[\bigl((t_2/t_1+1)^{2\alpha} -1)
\frac{\alpha \Gamma^2(\alpha)}{\Gamma(2\alpha+1)} - (t_2/t_1+1)^{2\alpha} \frac{(t_2/(t_1+t_2))^{\alpha+1}}{\alpha+1}
~_2F_1(1-\alpha,1+\alpha; \alpha+2; t_2/(t_1+t_2))\biggr],
\end{align*} 
\end{widetext}
where $~_2F_1(a,b;c;z)$ is the Gauss Hypergeometric function.
We conclude that
\begin{equation}\label{eq:Q_formula1}
\begin{split}
 Q(t_1,t_2) =& \frac{(t_2+t_1)^{2\alpha} - t_1^{2\alpha}}{T^{2\alpha} \Gamma(2\alpha+1)}
 - \frac{(t_1+t_2)^{\alpha-1} t_2^{\alpha+1}}{T^{2\alpha} \, \alpha (\alpha+1) \Gamma^2(\alpha)} 
\\ & \times
~_2F_1(1-\alpha,1+\alpha; \alpha+2; t_2/(t_1+t_2)),
\end{split}
\end{equation}
or, equivalently,
\begin{equation} \label{eq:Q2}
\begin{split}
 Q(t_1,t_2) =& \frac{t_1^{\alpha} (t_1+t_2)^\alpha}{T^{2\alpha} \, \alpha^2 \Gamma^2(\alpha)} 
~ _2F_1(-\alpha,\alpha; \alpha+1; t_1/(t_1 + t_2)) 
\\
&- \frac{t_1^{2\alpha}}{T^{2\alpha} \Gamma(2\alpha+1)} \,.
\end{split}
\end{equation}
For $\alpha = 1$, we retrieve again $Q(t_1,t_2) = t_1 t_2/T^2$.
\subsection{Statistics of random modulations}
The expectation of random modulations can be deduced directly from Eq.~(\ref{eq:def_modulation_from_S}) and Eq.~(\ref{eq:expec_jump_number}) as
\begin{equation}
 \langle J(t)\rangle=\langle \N(t)\rangle-\langle \N(t-\delta)\rangle.
\end{equation}
For an exponential distribution we have
\begin{equation}
 \langle J(t)\rangle=\delta/T,
\end{equation}
i.e., the modulations have a constant expectation.

For power-law waiting times with $0<\alpha<1$, Eq.~(\ref{eq:mean_number_jump_powerlaw}) gives 
\begin{equation}\label{eq:expectation_modulation}
\langle J(t)\rangle=\frac{t^\alpha-(t-\delta)^\alpha}{T^\alpha\Gamma(\alpha+1)}.
\end{equation}
At large $t\gg \delta$, one gets
\begin{equation}\langle J(t)\rangle\sim\alpha\delta\frac{t^{\alpha-1}}{T^\alpha\Gamma(\alpha+1)},
\end{equation}
thus the mean number of jumps realized between $t-\delta$ and $t$ decays as a power law of time: the
particle is progressively freezing.

The second moment can be obtained as
\begin{align*}
\langle J^2(t)\rangle =&
\langle (\N(t)-\N(t-\delta))^2\rangle\\
=&\langle \left(\N_2(\delta)+\N_1(t-\delta)-\N_1(t-\delta)\right)^2\rangle\\
=&\langle (\N_2(\delta)+\N_1(t-\delta))^2\rangle
\\&
-2\langle \N_1(t-\delta)\N_2(\delta)\rangle-\langle {\N}_1^2(t-\delta)\rangle\\
=&\langle {\N}^2(t)\rangle-2\langle \N_1(t-\delta)\N_2(\delta)\rangle-\langle {\N}^2(t-\delta)\rangle,
\end{align*}
where $\langle \N_1(t-\delta)\N_2(\delta)\rangle=
Q(t-\delta,\delta)$, from which the second moment is
\begin{equation}
\begin{split}
& \langle J^2(t)\rangle = \frac{t^\alpha-(t-\delta)^\alpha}{T^\alpha \Gamma(\alpha+1)} 
 \\&
 + \frac{2t^{\alpha-1}\delta^{\alpha+1}}{T^{2\alpha} \, \alpha(\alpha+1) \Gamma^2(\alpha)} 
~ _2F_1(1-\alpha,1+\alpha; \alpha+2; \delta/t),
\end{split}
\end{equation}
and the variance is
\begin{equation}
\begin{split}
 & \var (J(t)) =\\
 & \frac{t^\alpha-(t-\delta)^\alpha}{T^\alpha \Gamma(\alpha+1)}
 -\left(\frac{t^\alpha-(t-\delta)^\alpha}{T^\alpha \Gamma(\alpha+1)}\right)^2 
 \\&
 + \frac{2t^{\alpha-1}\delta^{\alpha+1}}{T^{2\alpha} \, \alpha(\alpha+1) \Gamma^2(\alpha)} 
~ _2F_1(1-\alpha,1+\alpha; \alpha+2; \delta/t),
\end{split}
\end{equation}
which decays as $t^{\alpha-1}$, just like the expectation.

We now consider the two-point average of having $n_1$ steps in the interval $(0,t_1)$ and $n$ steps in the interval $(0,t)$ where $t=t_1+t_2$ and $n=n_1+n_2$:
\begin{align} \nonumber
 \langle \N(t_1) \N(t)\rangle&=\langle \N_1(t_1) (\N_1(t_1)+\N_2(t_2) ) \rangle\\ \nonumber
 &=\langle \N_1^2(t_1)\rangle+ Q(t_1,t_2) \\
 &=\langle \N_1^2(t_1)\rangle+ Q(t_1,t-t_1).
\end{align}
As a consequence, we deduce
\begin{align} \label{eq:cov_operational_time}
\langle \N(t_1) \N(t)\rangle =& \frac{t_1^\alpha}{T^\alpha \Gamma(\alpha+1)} + \frac{t_1^{2\alpha}}{T^{2\alpha} \Gamma(2\alpha+1)}
\\\nonumber &+ \frac{t_1^{\alpha} t^\alpha}{T^{2\alpha} \, \alpha^2 \Gamma^2(\alpha)} 
~_2F_1(-\alpha,\alpha; \alpha+1; t_1/t).
\end{align}
To obtain the two-point average of random modulations, we average the product of increments during $\delta$
\begin{equation}\label{eq:prod_mom_modulation}
\begin{split}
 \langle J(t_1)J(t)\rangle=&\langle (\N(t_1)-\N(t_1-\delta))( \N(t)-\N(t-\delta))\rangle\\
 =& \langle \N(t_1) \N(t)\rangle - \langle \N(t_1)\N(t-\delta)\rangle
 \\&
 - \langle \N(t_1-\delta) \N(t)\rangle +\langle \N(t_1-\delta)\N(t-\delta)\rangle. 
\end{split}
\end{equation}
Combining Eq.~(\ref{eq:prod_mom_modulation}) and Eq.~(\ref{eq:expectation_modulation}) we finally obtain the covariance
\begin{widetext}
\begin{equation}
\begin{split}
&\cov(J(t_1),J(t))=
\frac{t_1^{\alpha} t^\alpha}{T^{2\alpha} \, \alpha^2 \Gamma^2(\alpha)} 
(~ _2F_1(-\alpha,\alpha; \alpha+1; t_1/t)-1)
-
\frac{(t_1-\delta)^{\alpha} t^\alpha}{T^{2\alpha} \, \alpha^2 \Gamma^2(\alpha)} 
(~ _2F_1(-\alpha,\alpha; \alpha+1; (t_1-\delta)/t)-1)
\\ &
-\frac{t_1^{\alpha} (t-\delta)^\alpha}{T^{2\alpha} \, \alpha^2 \Gamma^2(\alpha)} 
(~ _2F_1(-\alpha,\alpha; \alpha+1; t_1/(t-\delta))-1)
+
\frac{(t_1-\delta)^{\alpha} (t-\delta)^\alpha}{T^{2\alpha} \, \alpha^2 \Gamma^2(\alpha)} 
(~ _2F_1(-\alpha,\alpha; \alpha+1; (t_1-\delta)/(t-\delta))-1),
 \end{split}
 \end{equation}
\end{widetext}
which, for small $\delta\ll t_1$, amounts to the double derivative of the last term in Eq.~(\ref{eq:cov_operational_time}):
\begin{equation}
\begin{split}
\cov(J(t_1),J(t))\approx &\frac{\partial^2}{\partial t_1\partial t}\left[ \frac{t_1^{\alpha} t^\alpha}{T^{2\alpha} \, \alpha^2 \Gamma^2(\alpha)} \right.
\\
&\times ~ _2F_1(-\alpha,\alpha; \alpha+1; t_1/t)\Big].
\end{split}
\end{equation}
The covariance of random modulations for CTRW-Pow is not stationary.

In turn, the covariance for exponential waiting times (Markovian case) reads
\begin{equation}
 \cov(J(t_1),J(t))=
 \left\lbrace
 \begin{array}{ll}
 \frac{\delta^2}{T^2} & \mathrm{ for }\,\, t_1=t  \\
0  &  \mathrm{ otherwise } 
 \end{array}
\right.
\end{equation}
 which is a signature of an uncorrelated process. Therefore we conclude that random modulations for CTRW with exponential waiting times are random but uncorrelated; they constitute the simplest example of random modulations.



\section{Random modulations as a squared Gaussian process}
\label{appendix:square_gaussian_process}

In this Section, we compute exactly the expectation and the
covariance matrix of random modulations when they stem from the square
of a general Gaussian process. A notable example is the square of an
Ornstein-Uhlenbeck process, which is a minimal model for fluctuating diffusivity with tunable correlations ~\cite{PhysRevX.7.021002,sposini_etal-njp-2018}. However, our result is much wider and allows for a variety of covariance structures, especially the stationary ones. Since the covariance structure of random modulations determines the rate of convergence to the Gaussian distribution (see Eq.~(\ref{eq:NG_param}) and Sec.~\ref{sec:special_cases_NG_param}), it is useful to have access to a wide range of covariance structures including exponential, Gaussian, Mat\'ern, or even rational quadratic. 
A general Gaussian vector can be written as
\begin{equation}
 Y=\mu+\sqrt{Q}\eta,
\end{equation}
where $\mu$ is an expectation vector, $Q$ is a covariance matrix, and $\eta$ is the vector of standard IID normal variables.
The vector of random modulations is postulated to be the diagonal vector of the matrix $YY^\top$:
\begin{equation}
 J=\diag(YY^\top).
\end{equation}
The expectation of the matrix is
\begin{equation}
\langle YY^\top\rangle=\mu\mu^\top+Q,
\end{equation}
from which the mean random modulations are
\begin{equation}
 \langle J_i\rangle=\mu_i^2+Q_{ii}.
\end{equation}
The covariance matrix of random modulations can also be computed 
exactly. Setting $K=\sqrt{Q}$, one gets
\begin{align}\nonumber
 \cov(J_i,J_j)&=4\mu_i\mu_j \sum_{k=1}^NK_{ik}K_{jk}+2\left(\sum_{k=1}^NK_{ik}K_{jk}\right)^2 \\
 &=4\mu_iQ_{ij}\mu_j+2Q_{ij}^2.
\end{align}

The minimal model for diffusing diffusivity at equilibrium is obtained by choosing a constant mean $\mu_i=\mu$ and the covariance matrix of an OU process. The $d$-dimensional case is also accessible by setting $Y$ to be a $N\times d$ matrix.



\section{Simulation details}
\label{appendix:simulation details}
\subsection{Continuous-time random walk (CTRW)}

We consider a CTRW with a Gaussian jump PDF characterized by a diffusivity $D$ and the Mittag-Leffler distributed waiting times with the parameter $\alpha$ and timescale $T$ (see Appendix~\ref{appendix:CTRW}). The CTRW has been simulated by taking a diagonal covariance matrix with elements $C_{ii}=D\delta$.
 The matrix $J$ is constructed by generating a renewal process with randomly distributed waiting times. Then each modulation $J_i$ is defined as the number of renewal events occurred between timestep $(i-1)\delta$ and $i\delta$.
 {\clr We distinguish two cases: (i) CTRW-Pow model where the waiting time distribution is Mittag-Leffler with parameter $\alpha$ (Appendix~\ref{sec:expec_var_tot_num_jump}); other parameters are $D=1$, $\alpha=0.7$, $T=1$, and $\delta=1$; and CTRW-Exp where the waiting-time distribution is exponential, with parameters $D=1$, $T=1$, and $\delta=1$. }
\subsection{Fractional Brownian motion (fBm)}
We consider a fBm with generalized diffusion coefficient $D$ and Hurst exponent $H$ related to the anomalous diffusion exponent through $\alpha=2H$. The fBm is obtained by setting $C$ to be the covariance matrix of the fractional Gaussian noise
\begin{equation}\label{eq:C_for_fBm}
\begin{split}    
 &C(i,j)=
 \\
 &
 D\delta^{2H}\left(|i-j-1|^{2H}-2|i-j|^{2H}+|i-j+1|^{2H}\right).
\end{split}
\end{equation}
The Cholesky decomposition of $C$ is performed to obtain the square root $\sqrt{C}$ in the form of a lower triangular matrix verifying $C=\sqrt{C}\sqrt{C}^\top$. The modulation matrix is the identity: $J=I$. The chosen parameters are $D=1$, $\delta=1$, and $H=0.35$.

\subsection{Scaled Brownian motion (sBm)}
We also model sBm, a Markovian process with time dependent diffusivity $D(i)=D_0\alpha i^{\alpha-1}$, where $D_0$ is the initial diffusion coefficient. 
For sBm, the increments are independent so that the covariance matrix is diagonal with the elements $C_{ii}=2D_0\delta$. The modulations mimic the progressive cooling of the system through the deterministic relation $J_{ii}=\alpha i^{\alpha-1}$. The chosen parameter are $\alpha=0.7$, $D_0=1$, and $\delta=1$.

\subsection{Switching Diffusivity (SD-Exp)}
Diffusion with switching diffusivity alternates between two diffusion coefficients $D_1$ and $D_2$ with prescribed switching rates. SD-Exp is simulated by using a diagonal covariance matrix $C_{ii}=2D_1\delta$, the modulations alternate between two values. The diagonal elements $J_{ii}$ are generated by a Markov chain dynamics, with a probability $p_1$ to remain in state $1$ with $J_i=1$ and probability $p_2$ to remain in state $2$ with $J_i=D_2/D_1$. Initial state is chosen according to the equilibrium probabilities $p_1^{\textrm{(eq)}}=p_1/(p_1+p_2)$ and $p_2^{\textrm{(eq)}}=p_2/(p_1+p_2)$. The chosen parameters are $D_1=1$, $D_2=0$, and $p_1=p_2=0.995$.

\subsection{Diffusing diffusivity (DD-Exp)}
\label{sec:sim_DD-Exp}
For DD-Exp we model the diagonal covariance matrix
$C_{ii}=2\bar{D}\delta$, where $\bar{D}$ is the equilibrium diffusion coefficient. The random modulations are obtained by rescaling the Cox-Ingersoll-Ross (CIR) process from~\cite{Lanoiselee2018_NGmodel,Lanoiselee2018_FPT} by $\bar{D}$, from which $J_t$ follows 
\begin{equation}
dJ_t=\frac{1}{\tau}(1-J_t)dt+\sqrt{\frac{2J_t}{\nu}}dW_t,
\end{equation}
where $\tau$ is the correlation time of random modulations, $\nu$ controls the strength of fluctuations, and $dW_t$ are the increments of the Wiener process. The diagonal elements of the random modulation matrix are obtained by simulation of the CIR process.
The simulation of CIR does not rely on an approximation scheme but takes advantage of the exact knowledge of the transition probabilities. The simulation is performed as follows:
The first modulation $J_1$ is drawn from a Gamma distribution with the shape parameter $\nu$ and the scale equal to $1$:
\begin{equation}
 p(J_1)=\frac{1}{\Gamma(\nu)}J_1^{\nu-1}e^{-J_1}.
\end{equation}
The subsequent modulations are obtained iteratively
\begin{equation}
 J_{n+1}=J_n+\eta(J_n,\delta)/(2c),
\end{equation}
where $c=\tau(1-e^{-\delta/\tau})$ and $\eta(J_n,\delta)$ is a non-central $\chi^2$ random variable with the degree of freedom $2\nu$ and the non-centrality parameter $2cJ_ne^{-\delta/\tau}$~\cite{Cox1985}.

The parameters chosen for simulations are $\bar{D}=1$, $\nu=1$ and $\delta=1$.

{\clr
\subsection{Random Constant model (Rand-Const)}
Finally, we simulate processes with random but trajectory-wise constant modulations by taking, for a given trajectory $m$, $C_{ii}=D\delta$ and constant modulations $J_i=\mu^{(m)}$ where $\mu^{(m)}$ is a random variable drawn from a given distribution for each $m$. In the simulations, the chosen parameters are $\delta=1$, $D=1/2$, and $\mu^{(m)}$ exponentially distributed with mean $\langle \mu^{(m)}\rangle=1 $.}

\bibliographystyle{apsrev4-2} 

\begin{thebibliography}{89}%
\makeatletter
\providecommand \@ifxundefined [1]{%
 \@ifx{#1\undefined}
}%
\providecommand \@ifnum [1]{%
 \ifnum #1\expandafter \@firstoftwo
 \else \expandafter \@secondoftwo
 \fi
}%
\providecommand \@ifx [1]{%
 \ifx #1\expandafter \@firstoftwo
 \else \expandafter \@secondoftwo
 \fi
}%
\providecommand \natexlab [1]{#1}%
\providecommand \enquote  [1]{``#1''}%
\providecommand \bibnamefont  [1]{#1}%
\providecommand \bibfnamefont [1]{#1}%
\providecommand \citenamefont [1]{#1}%
\providecommand \href@noop [0]{\@secondoftwo}%
\providecommand \href [0]{\begingroup \@sanitize@url \@href}%
\providecommand \@href[1]{\@@startlink{#1}\@@href}%
\providecommand \@@href[1]{\endgroup#1\@@endlink}%
\providecommand \@sanitize@url [0]{\catcode `\\12\catcode `\$12\catcode
  `\&12\catcode `\#12\catcode `\^12\catcode `\_12\catcode `\%12\relax}%
\providecommand \@@startlink[1]{}%
\providecommand \@@endlink[0]{}%
\providecommand \url  [0]{\begingroup\@sanitize@url \@url }%
\providecommand \@url [1]{\endgroup\@href {#1}{\urlprefix }}%
\providecommand \urlprefix  [0]{URL }%
\providecommand \Eprint [0]{\href }%
\providecommand \doibase [0]{https://doi.org/}%
\providecommand \selectlanguage [0]{\@gobble}%
\providecommand \bibinfo  [0]{\@secondoftwo}%
\providecommand \bibfield  [0]{\@secondoftwo}%
\providecommand \translation [1]{[#1]}%
\providecommand \BibitemOpen [0]{}%
\providecommand \bibitemStop [0]{}%
\providecommand \bibitemNoStop [0]{.\EOS\space}%
\providecommand \EOS [0]{\spacefactor3000\relax}%
\providecommand \BibitemShut  [1]{\csname bibitem#1\endcsname}%
\let\auto@bib@innerbib\@empty
\bibitem [{\citenamefont {Pogany}(1976)}]{POGANY1976}%
  \BibitemOpen
  \bibfield  {author} {\bibinfo {author} {\bibfnamefont {G.}~\bibnamefont
  {Pogany}},\ }\href
  {https://doi.org/https://doi.org/10.1016/0032-3861(76)90209-3} {\bibfield
  {journal} {\bibinfo  {journal} {Polymer}\ }\textbf {\bibinfo {volume} {17}},\
  \bibinfo {pages} {690} (\bibinfo {year} {1976})}\BibitemShut {NoStop}%
\bibitem [{\citenamefont {Stariolo}\ and\ \citenamefont
  {Fabricius}(2006)}]{Stariolo2006}%
  \BibitemOpen
  \bibfield  {author} {\bibinfo {author} {\bibfnamefont {D.~A.}\ \bibnamefont
  {Stariolo}}\ and\ \bibinfo {author} {\bibfnamefont {G.}~\bibnamefont
  {Fabricius}},\ }\href {https://doi.org/10.1063/1.2221309} {\bibfield
  {journal} {\bibinfo  {journal} {J. Chem. Phys.}\ }\textbf {\bibinfo {volume}
  {125}},\ \bibinfo {pages} {064505} (\bibinfo {year} {2006})}\BibitemShut
  {NoStop}%
\bibitem [{\citenamefont {Chaudhuri}\ \emph {et~al.}(2007)\citenamefont
  {Chaudhuri}, \citenamefont {Berthier},\ and\ \citenamefont
  {Kob}}]{Chaudhuri2007}%
  \BibitemOpen
  \bibfield  {author} {\bibinfo {author} {\bibfnamefont {P.}~\bibnamefont
  {Chaudhuri}}, \bibinfo {author} {\bibfnamefont {L.}~\bibnamefont
  {Berthier}},\ and\ \bibinfo {author} {\bibfnamefont {W.}~\bibnamefont
  {Kob}},\ }\href {https://doi.org/10.1103/PhysRevLett.99.060604} {\bibfield
  {journal} {\bibinfo  {journal} {Phys. Rev. Lett.}\ }\textbf {\bibinfo
  {volume} {99}},\ \bibinfo {pages} {060604} (\bibinfo {year}
  {2007})}\BibitemShut {NoStop}%
\bibitem [{\citenamefont {Sagi}\ \emph {et~al.}(2012)\citenamefont {Sagi},
  \citenamefont {Brook}, \citenamefont {Almog},\ and\ \citenamefont
  {Davidson}}]{Sagi2012}%
  \BibitemOpen
  \bibfield  {author} {\bibinfo {author} {\bibfnamefont {Y.}~\bibnamefont
  {Sagi}}, \bibinfo {author} {\bibfnamefont {M.}~\bibnamefont {Brook}},
  \bibinfo {author} {\bibfnamefont {I.}~\bibnamefont {Almog}},\ and\ \bibinfo
  {author} {\bibfnamefont {N.}~\bibnamefont {Davidson}},\ }\href
  {https://doi.org/10.1103/PhysRevLett.108.093002} {\bibfield  {journal}
  {\bibinfo  {journal} {Phys. Rev. Lett.}\ }\textbf {\bibinfo {volume} {108}},\
  \bibinfo {pages} {093002} (\bibinfo {year} {2012})}\BibitemShut {NoStop}%
\bibitem [{\citenamefont {Jeon}\ \emph {et~al.}(2013)\citenamefont {Jeon},
  \citenamefont {Leijnse}, \citenamefont {Oddershede},\ and\ \citenamefont
  {Metzler}}]{Jeon2013}%
  \BibitemOpen
  \bibfield  {author} {\bibinfo {author} {\bibfnamefont {J.-H.}\ \bibnamefont
  {Jeon}}, \bibinfo {author} {\bibfnamefont {N.}~\bibnamefont {Leijnse}},
  \bibinfo {author} {\bibfnamefont {L.~B.}\ \bibnamefont {Oddershede}},\ and\
  \bibinfo {author} {\bibfnamefont {R.}~\bibnamefont {Metzler}},\ }\href
  {https://doi.org/10.1088/1367-2630/15/4/045011} {\bibfield  {journal}
  {\bibinfo  {journal} {New J. Phys.}\ }\textbf {\bibinfo {volume} {15}},\
  \bibinfo {pages} {045011} (\bibinfo {year} {2013})}\BibitemShut {NoStop}%
\bibitem [{\citenamefont {Leptos}\ \emph {et~al.}(2009)\citenamefont {Leptos},
  \citenamefont {Guasto}, \citenamefont {Gollub}, \citenamefont {Pesci},\ and\
  \citenamefont {Goldstein}}]{Leptos2009}%
  \BibitemOpen
  \bibfield  {author} {\bibinfo {author} {\bibfnamefont {K.~C.}\ \bibnamefont
  {Leptos}}, \bibinfo {author} {\bibfnamefont {J.~S.}\ \bibnamefont {Guasto}},
  \bibinfo {author} {\bibfnamefont {J.~P.}\ \bibnamefont {Gollub}}, \bibinfo
  {author} {\bibfnamefont {A.~I.}\ \bibnamefont {Pesci}},\ and\ \bibinfo
  {author} {\bibfnamefont {R.~E.}\ \bibnamefont {Goldstein}},\ }\href
  {https://doi.org/10.1103/PhysRevLett.103.198103} {\bibfield  {journal}
  {\bibinfo  {journal} {Phys. Rev. Lett.}\ }\textbf {\bibinfo {volume} {103}},\
  \bibinfo {pages} {198103} (\bibinfo {year} {2009})}\BibitemShut {NoStop}%
\bibitem [{\citenamefont {Chakraborty}\ and\ \citenamefont
  {Roichman}(2020)}]{Chakraborty2020}%
  \BibitemOpen
  \bibfield  {author} {\bibinfo {author} {\bibfnamefont {I.}~\bibnamefont
  {Chakraborty}}\ and\ \bibinfo {author} {\bibfnamefont {Y.}~\bibnamefont
  {Roichman}},\ }\href {https://doi.org/10.1103/PhysRevResearch.2.022020}
  {\bibfield  {journal} {\bibinfo  {journal} {Phys. Rev. Res.}\ }\textbf
  {\bibinfo {volume} {2}},\ \bibinfo {pages} {022020} (\bibinfo {year}
  {2020})}\BibitemShut {NoStop}%
\bibitem [{\citenamefont {Bronstein}\ \emph {et~al.}(2009)\citenamefont
  {Bronstein}, \citenamefont {Israel}, \citenamefont {Kepten}, \citenamefont
  {Mai}, \citenamefont {Shav-Tal}, \citenamefont {Barkai},\ and\ \citenamefont
  {Garini}}]{Bronstein2009}%
  \BibitemOpen
  \bibfield  {author} {\bibinfo {author} {\bibfnamefont {I.}~\bibnamefont
  {Bronstein}}, \bibinfo {author} {\bibfnamefont {Y.}~\bibnamefont {Israel}},
  \bibinfo {author} {\bibfnamefont {E.}~\bibnamefont {Kepten}}, \bibinfo
  {author} {\bibfnamefont {S.}~\bibnamefont {Mai}}, \bibinfo {author}
  {\bibfnamefont {Y.}~\bibnamefont {Shav-Tal}}, \bibinfo {author}
  {\bibfnamefont {E.}~\bibnamefont {Barkai}},\ and\ \bibinfo {author}
  {\bibfnamefont {Y.}~\bibnamefont {Garini}},\ }\href
  {https://doi.org/10.1103/PhysRevLett.103.018102} {\bibfield  {journal}
  {\bibinfo  {journal} {Phys. Rev. Lett.}\ }\textbf {\bibinfo {volume} {103}},\
  \bibinfo {pages} {018102} (\bibinfo {year} {2009})}\BibitemShut {NoStop}%
\bibitem [{\citenamefont {Witzel}\ \emph {et~al.}(2019)\citenamefont {Witzel},
  \citenamefont {G{\"o}tz}, \citenamefont {Lanoisel{\'e}e}, \citenamefont
  {Franosch}, \citenamefont {Grebenkov},\ and\ \citenamefont
  {Heinrich}}]{Witzel2019}%
  \BibitemOpen
  \bibfield  {author} {\bibinfo {author} {\bibfnamefont {P.}~\bibnamefont
  {Witzel}}, \bibinfo {author} {\bibfnamefont {M.}~\bibnamefont {G{\"o}tz}},
  \bibinfo {author} {\bibfnamefont {Y.}~\bibnamefont {Lanoisel{\'e}e}},
  \bibinfo {author} {\bibfnamefont {T.}~\bibnamefont {Franosch}}, \bibinfo
  {author} {\bibfnamefont {D.~S.}\ \bibnamefont {Grebenkov}},\ and\ \bibinfo
  {author} {\bibfnamefont {D.}~\bibnamefont {Heinrich}},\ }\href
  {https://doi.org/10.1016/j.bpj.2019.06.009} {\bibfield  {journal} {\bibinfo
  {journal} {Biophys. J.}\ }\textbf {\bibinfo {volume} {117}},\ \bibinfo
  {pages} {203} (\bibinfo {year} {2019})}\BibitemShut {NoStop}%
\bibitem [{\citenamefont {Sabri}\ \emph {et~al.}(2020)\citenamefont {Sabri},
  \citenamefont {Xu}, \citenamefont {Krapf},\ and\ \citenamefont
  {Weiss}}]{Sabri2020}%
  \BibitemOpen
  \bibfield  {author} {\bibinfo {author} {\bibfnamefont {A.}~\bibnamefont
  {Sabri}}, \bibinfo {author} {\bibfnamefont {X.}~\bibnamefont {Xu}}, \bibinfo
  {author} {\bibfnamefont {D.}~\bibnamefont {Krapf}},\ and\ \bibinfo {author}
  {\bibfnamefont {M.}~\bibnamefont {Weiss}},\ }\href
  {https://doi.org/10.1103/PhysRevLett.125.058101} {\bibfield  {journal}
  {\bibinfo  {journal} {Phys. Rev. Lett.}\ }\textbf {\bibinfo {volume} {125}},\
  \bibinfo {pages} {058101} (\bibinfo {year} {2020})}\BibitemShut {NoStop}%
\bibitem [{\citenamefont {Wang}\ \emph {et~al.}(2009)\citenamefont {Wang},
  \citenamefont {Anthony}, \citenamefont {Bae},\ and\ \citenamefont
  {Granick}}]{Wang2009}%
  \BibitemOpen
  \bibfield  {author} {\bibinfo {author} {\bibfnamefont {B.}~\bibnamefont
  {Wang}}, \bibinfo {author} {\bibfnamefont {S.~M.}\ \bibnamefont {Anthony}},
  \bibinfo {author} {\bibfnamefont {S.~C.}\ \bibnamefont {Bae}},\ and\ \bibinfo
  {author} {\bibfnamefont {S.}~\bibnamefont {Granick}},\ }\href
  {https://doi.org/10.1073/pnas.0903554106} {\bibfield  {journal} {\bibinfo
  {journal} {Proc. Nat. Acad. Sci.}\ }\textbf {\bibinfo {volume} {106}},\
  \bibinfo {pages} {15160} (\bibinfo {year} {2009})}\BibitemShut {NoStop}%
\bibitem [{\citenamefont {Weigel}\ \emph {et~al.}(2011)\citenamefont {Weigel},
  \citenamefont {Simon}, \citenamefont {Tamkun},\ and\ \citenamefont
  {Krapf}}]{Weigel2011}%
  \BibitemOpen
  \bibfield  {author} {\bibinfo {author} {\bibfnamefont {A.~V.}\ \bibnamefont
  {Weigel}}, \bibinfo {author} {\bibfnamefont {B.}~\bibnamefont {Simon}},
  \bibinfo {author} {\bibfnamefont {M.~M.}\ \bibnamefont {Tamkun}},\ and\
  \bibinfo {author} {\bibfnamefont {D.}~\bibnamefont {Krapf}},\ }\href
  {https://doi.org/10.1073/pnas.1016325108} {\bibfield  {journal} {\bibinfo
  {journal} {Proc. Nat. Acad. Sci.}\ }\textbf {\bibinfo {volume} {108}},\
  \bibinfo {pages} {6438} (\bibinfo {year} {2011})}\BibitemShut {NoStop}%
\bibitem [{\citenamefont {Sungkaworn}\ \emph {et~al.}(2017)\citenamefont
  {Sungkaworn}, \citenamefont {Jobin}, \citenamefont {Burnecki}, \citenamefont
  {Weron}, \citenamefont {Lohse},\ and\ \citenamefont
  {Calebiro}}]{Sungkaworn2017-db}%
  \BibitemOpen
  \bibfield  {author} {\bibinfo {author} {\bibfnamefont {T.}~\bibnamefont
  {Sungkaworn}}, \bibinfo {author} {\bibfnamefont {M.-L.}\ \bibnamefont
  {Jobin}}, \bibinfo {author} {\bibfnamefont {K.}~\bibnamefont {Burnecki}},
  \bibinfo {author} {\bibfnamefont {A.}~\bibnamefont {Weron}}, \bibinfo
  {author} {\bibfnamefont {M.~J.}\ \bibnamefont {Lohse}},\ and\ \bibinfo
  {author} {\bibfnamefont {D.}~\bibnamefont {Calebiro}},\ }\href
  {https://doi.org/https://doi.org/10.1038/nature24264} {\bibfield  {journal}
  {\bibinfo  {journal} {Nature}\ }\textbf {\bibinfo {volume} {550}},\ \bibinfo
  {pages} {543} (\bibinfo {year} {2017})}\BibitemShut {NoStop}%
\bibitem [{\citenamefont {Grimes}\ \emph {et~al.}(2023)\citenamefont {Grimes},
  \citenamefont {Koszegi}, \citenamefont {Lanoisel{\'e}e}, \citenamefont
  {Miljus}, \citenamefont {O'Brien}, \citenamefont {Stepniewski}, \citenamefont
  {Medel-Lacruz}, \citenamefont {Baidya}, \citenamefont {Makarova},
  \citenamefont {Mistry}, \citenamefont {Goulding}, \citenamefont {Drube},
  \citenamefont {Hoffmann}, \citenamefont {Owen}, \citenamefont {Shukla},
  \citenamefont {Selent}, \citenamefont {Hill},\ and\ \citenamefont
  {Calebiro}}]{Grimes2023-ns}%
  \BibitemOpen
  \bibfield  {author} {\bibinfo {author} {\bibfnamefont {J.}~\bibnamefont
  {Grimes}}, \bibinfo {author} {\bibfnamefont {Z.}~\bibnamefont {Koszegi}},
  \bibinfo {author} {\bibfnamefont {Y.}~\bibnamefont {Lanoisel{\'e}e}},
  \bibinfo {author} {\bibfnamefont {T.}~\bibnamefont {Miljus}}, \bibinfo
  {author} {\bibfnamefont {S.~L.}\ \bibnamefont {O'Brien}}, \bibinfo {author}
  {\bibfnamefont {T.~M.}\ \bibnamefont {Stepniewski}}, \bibinfo {author}
  {\bibfnamefont {B.}~\bibnamefont {Medel-Lacruz}}, \bibinfo {author}
  {\bibfnamefont {M.}~\bibnamefont {Baidya}}, \bibinfo {author} {\bibfnamefont
  {M.}~\bibnamefont {Makarova}}, \bibinfo {author} {\bibfnamefont
  {R.}~\bibnamefont {Mistry}}, \bibinfo {author} {\bibfnamefont
  {J.}~\bibnamefont {Goulding}}, \bibinfo {author} {\bibfnamefont
  {J.}~\bibnamefont {Drube}}, \bibinfo {author} {\bibfnamefont
  {C.}~\bibnamefont {Hoffmann}}, \bibinfo {author} {\bibfnamefont {D.~M.}\
  \bibnamefont {Owen}}, \bibinfo {author} {\bibfnamefont {A.~K.}\ \bibnamefont
  {Shukla}}, \bibinfo {author} {\bibfnamefont {J.}~\bibnamefont {Selent}},
  \bibinfo {author} {\bibfnamefont {S.~J.}\ \bibnamefont {Hill}},\ and\
  \bibinfo {author} {\bibfnamefont {D.}~\bibnamefont {Calebiro}},\ }\href
  {https://doi.org/10.1016/j.cell.2023.04.018} {\bibfield  {journal} {\bibinfo
  {journal} {Cell}\ }\textbf {\bibinfo {volume} {186}},\ \bibinfo {pages}
  {2238} (\bibinfo {year} {2023})}\BibitemShut {NoStop}%
\bibitem [{\citenamefont {Ben-Avraham}\ and\ \citenamefont
  {Havlin}(2000)}]{Ben-Avraham}%
  \BibitemOpen
  \bibfield  {author} {\bibinfo {author} {\bibfnamefont {D.}~\bibnamefont
  {Ben-Avraham}}\ and\ \bibinfo {author} {\bibfnamefont {S.}~\bibnamefont
  {Havlin}},\ }\href@noop {} {\emph {\bibinfo {title} {Diffusion and reaction
  in disordered systems}}}\ (\bibinfo  {publisher} {Cambridge University
  Press},\ \bibinfo {year} {2000})\BibitemShut {NoStop}%
\bibitem [{\citenamefont {Metzler}\ \emph {et~al.}(2014)\citenamefont
  {Metzler}, \citenamefont {Oshanin},\ and\ \citenamefont {Redner}}]{Metzler}%
  \BibitemOpen
  \bibfield  {author} {\bibinfo {author} {\bibfnamefont {R.}~\bibnamefont
  {Metzler}}, \bibinfo {author} {\bibfnamefont {G.}~\bibnamefont {Oshanin}},\
  and\ \bibinfo {author} {\bibfnamefont {S.}~\bibnamefont {Redner}},\
  }\href@noop {} {\emph {\bibinfo {title} {First-Passage Phenomena and Their
  Applications}}}\ (\bibinfo  {publisher} {World Scientific, Singapore},\
  \bibinfo {year} {2014})\BibitemShut {NoStop}%
\bibitem [{\citenamefont {Lindenberg}\ \emph {et~al.}(2019)\citenamefont
  {Lindenberg}, \citenamefont {Metzler},\ and\ \citenamefont
  {Oshanin}}]{Lindenberg}%
  \BibitemOpen
  \bibfield  {author} {\bibinfo {author} {\bibfnamefont {K.}~\bibnamefont
  {Lindenberg}}, \bibinfo {author} {\bibfnamefont {R.}~\bibnamefont
  {Metzler}},\ and\ \bibinfo {author} {\bibfnamefont {G.}~\bibnamefont
  {Oshanin}},\ }\href@noop {} {\emph {\bibinfo {title} {Chemical Kinetics:
  Beyond the Textbook}}}\ (\bibinfo  {publisher} {World Scientific, New
  Jersey},\ \bibinfo {year} {2019})\BibitemShut {NoStop}%
\bibitem [{\citenamefont {Grebenkov}\ \emph {et~al.}(2024)\citenamefont
  {Grebenkov}, \citenamefont {Metzler},\ and\ \citenamefont
  {Oshanin}}]{Grebenkov}%
  \BibitemOpen
  \bibfield  {author} {\bibinfo {author} {\bibfnamefont {D.~S.}\ \bibnamefont
  {Grebenkov}}, \bibinfo {author} {\bibfnamefont {R.}~\bibnamefont {Metzler}},\
  and\ \bibinfo {author} {\bibfnamefont {G.}~\bibnamefont {Oshanin}},\
  }\href@noop {} {\emph {\bibinfo {title} {Target Search Problems}}}\ (\bibinfo
   {publisher} {Springer: Cham, Switzerland},\ \bibinfo {year}
  {2024})\BibitemShut {NoStop}%
\bibitem [{\citenamefont {Metzler}\ and\ \citenamefont
  {Klafter}(2000)}]{METZLER2000}%
  \BibitemOpen
  \bibfield  {author} {\bibinfo {author} {\bibfnamefont {R.}~\bibnamefont
  {Metzler}}\ and\ \bibinfo {author} {\bibfnamefont {J.}~\bibnamefont
  {Klafter}},\ }\href
  {https://doi.org/https://doi.org/10.1016/S0370-1573(00)00070-3} {\bibfield
  {journal} {\bibinfo  {journal} {Phys. Rep.}\ }\textbf {\bibinfo {volume}
  {339}},\ \bibinfo {pages} {1} (\bibinfo {year} {2000})}\BibitemShut {NoStop}%
\bibitem [{\citenamefont {Kutner}\ and\ \citenamefont
  {Masoliver}(2017)}]{Kutner2017}%
  \BibitemOpen
  \bibfield  {author} {\bibinfo {author} {\bibfnamefont {R.}~\bibnamefont
  {Kutner}}\ and\ \bibinfo {author} {\bibfnamefont {J.}~\bibnamefont
  {Masoliver}},\ }\href {https://doi.org/10.1140/epjb/e2016-70578-3} {\bibfield
   {journal} {\bibinfo  {journal} {Euro. Phys. J. B}\ }\textbf {\bibinfo
  {volume} {90}},\ \bibinfo {pages} {50} (\bibinfo {year} {2017})}\BibitemShut
  {NoStop}%
\bibitem [{\citenamefont {Wyss}(1986)}]{wyss-jmp-1986}%
  \BibitemOpen
  \bibfield  {author} {\bibinfo {author} {\bibfnamefont {W.}~\bibnamefont
  {Wyss}},\ }\href {https://doi.org/https://doi.org/10.1063/1.527251}
  {\bibfield  {journal} {\bibinfo  {journal} {J. Math. Phys.}\ }\textbf
  {\bibinfo {volume} {27}},\ \bibinfo {pages} {2782} (\bibinfo {year}
  {1986})}\BibitemShut {NoStop}%
\bibitem [{\citenamefont {Hilfer}\ and\ \citenamefont
  {Anton}(1995)}]{hilfer_etal-pre-1995}%
  \BibitemOpen
  \bibfield  {author} {\bibinfo {author} {\bibfnamefont {R.}~\bibnamefont
  {Hilfer}}\ and\ \bibinfo {author} {\bibfnamefont {L.}~\bibnamefont {Anton}},\
  }\href {https://doi.org/https://doi.org/10.1103/PhysRevE.51.R848} {\bibfield
  {journal} {\bibinfo  {journal} {Phys. Rev. E}\ }\textbf {\bibinfo {volume}
  {51}},\ \bibinfo {pages} {R848} (\bibinfo {year} {1995})}\BibitemShut
  {NoStop}%
\bibitem [{\citenamefont {Mainardi}(1996)}]{mainardi-csf-1996}%
  \BibitemOpen
  \bibfield  {author} {\bibinfo {author} {\bibfnamefont {F.}~\bibnamefont
  {Mainardi}},\ }\href
  {https://doi.org/https://doi.org/10.1016/0960-0779(95)00125-5} {\bibfield
  {journal} {\bibinfo  {journal} {Chaos Solit. Fract.}\ }\textbf {\bibinfo
  {volume} {7}},\ \bibinfo {pages} {1461} (\bibinfo {year} {1996})}\BibitemShut
  {NoStop}%
\bibitem [{\citenamefont {Jeon}\ \emph {et~al.}(2014)\citenamefont {Jeon},
  \citenamefont {Chechkin},\ and\ \citenamefont {Metzler}}]{Jeon2014}%
  \BibitemOpen
  \bibfield  {author} {\bibinfo {author} {\bibfnamefont {J.-H.}\ \bibnamefont
  {Jeon}}, \bibinfo {author} {\bibfnamefont {A.~V.}\ \bibnamefont {Chechkin}},\
  and\ \bibinfo {author} {\bibfnamefont {R.}~\bibnamefont {Metzler}},\ }\href
  {https://doi.org/10.1039/c4cp02019g} {\bibfield  {journal} {\bibinfo
  {journal} {Physical Chemistry Chemical Physics}\ }\textbf {\bibinfo {volume}
  {16}},\ \bibinfo {pages} {15811} (\bibinfo {year} {2014})}\BibitemShut
  {NoStop}%
\bibitem [{\citenamefont {Mandelbrot}\ and\ \citenamefont
  {Van~Ness}(1968)}]{Mandelbrot1968}%
  \BibitemOpen
  \bibfield  {author} {\bibinfo {author} {\bibfnamefont {B.~B.}\ \bibnamefont
  {Mandelbrot}}\ and\ \bibinfo {author} {\bibfnamefont {J.~W.}\ \bibnamefont
  {Van~Ness}},\ }\href {https://doi.org/10.1137/1010093} {\bibfield  {journal}
  {\bibinfo  {journal} {SIAM Rev.}\ }\textbf {\bibinfo {volume} {10}},\
  \bibinfo {pages} {422} (\bibinfo {year} {1968})}\BibitemShut {NoStop}%
\bibitem [{\citenamefont {Benelli}\ and\ \citenamefont
  {Weiss}(2021)}]{Benelli_2021}%
  \BibitemOpen
  \bibfield  {author} {\bibinfo {author} {\bibfnamefont {R.}~\bibnamefont
  {Benelli}}\ and\ \bibinfo {author} {\bibfnamefont {M.}~\bibnamefont
  {Weiss}},\ }\href {https://doi.org/10.1088/1367-2630/ac0853} {\bibfield
  {journal} {\bibinfo  {journal} {New J. Phys.}\ }\textbf {\bibinfo {volume}
  {23}},\ \bibinfo {pages} {063072} (\bibinfo {year} {2021})}\BibitemShut
  {NoStop}%
\bibitem [{\citenamefont {Zwanzig}(1973)}]{Zwanzig1973}%
  \BibitemOpen
  \bibfield  {author} {\bibinfo {author} {\bibfnamefont {R.}~\bibnamefont
  {Zwanzig}},\ }\href {https://doi.org/10.1007/BF01008729} {\bibfield
  {journal} {\bibinfo  {journal} {J. Stat. Phys.}\ }\textbf {\bibinfo {volume}
  {9}},\ \bibinfo {pages} {215} (\bibinfo {year} {1973})}\BibitemShut {NoStop}%
\bibitem [{\citenamefont {Porr\'a}\ \emph {et~al.}(1996)\citenamefont
  {Porr\'a}, \citenamefont {Wang},\ and\ \citenamefont
  {Masoliver}}]{Porra1996}%
  \BibitemOpen
  \bibfield  {author} {\bibinfo {author} {\bibfnamefont {J.~M.}\ \bibnamefont
  {Porr\'a}}, \bibinfo {author} {\bibfnamefont {K.-G.}\ \bibnamefont {Wang}},\
  and\ \bibinfo {author} {\bibfnamefont {J.}~\bibnamefont {Masoliver}},\
  }\href@noop {} {\bibfield  {journal} {\bibinfo  {journal} {Phys. Rev. E}\
  }\textbf {\bibinfo {volume} {53}},\ \bibinfo {pages} {5872} (\bibinfo {year}
  {1996})}\BibitemShut {NoStop}%
\bibitem [{\citenamefont {Bertseva}\ \emph {et~al.}(2012)\citenamefont
  {Bertseva}, \citenamefont {Grebenkov}, \citenamefont {Schmidhauser},
  \citenamefont {Gribkova}, \citenamefont {Jeney},\ and\ \citenamefont
  {Forro}}]{Bertseva2012}%
  \BibitemOpen
  \bibfield  {author} {\bibinfo {author} {\bibfnamefont {E.}~\bibnamefont
  {Bertseva}}, \bibinfo {author} {\bibfnamefont {D.~S.}\ \bibnamefont
  {Grebenkov}}, \bibinfo {author} {\bibfnamefont {P.}~\bibnamefont
  {Schmidhauser}}, \bibinfo {author} {\bibfnamefont {S.}~\bibnamefont
  {Gribkova}}, \bibinfo {author} {\bibfnamefont {S.}~\bibnamefont {Jeney}},\
  and\ \bibinfo {author} {\bibfnamefont {L.}~\bibnamefont {Forro}},\ }\href
  {https://doi.org/10.1140/epje/i2012-12063-4} {\bibfield  {journal} {\bibinfo
  {journal} {Eur. Phys. J. E}\ }\textbf {\bibinfo {volume} {35}},\ \bibinfo
  {pages} {63} (\bibinfo {year} {2012})}\BibitemShut {NoStop}%
\bibitem [{\citenamefont {Goychuk}(2012)}]{Goychuk2012}%
  \BibitemOpen
  \bibfield  {author} {\bibinfo {author} {\bibfnamefont {I.}~\bibnamefont
  {Goychuk}},\ }\href
  {https://doi.org/https://doi.org/10.1002/9781118197714.ch5} {\bibfield
  {journal} {\bibinfo  {journal} {Adv. Chem. Phys.}\ }\textbf {\bibinfo
  {volume} {150}},\ \bibinfo {pages} {187} (\bibinfo {year}
  {2012})}\BibitemShut {NoStop}%
\bibitem [{\citenamefont {Fogedby}(1994)}]{Fogedby1994}%
  \BibitemOpen
  \bibfield  {author} {\bibinfo {author} {\bibfnamefont {H.~C.}\ \bibnamefont
  {Fogedby}},\ }\href {https://doi.org/10.1103/PhysRevLett.73.2517} {\bibfield
  {journal} {\bibinfo  {journal} {Phys. Rev. Lett.}\ }\textbf {\bibinfo
  {volume} {73}},\ \bibinfo {pages} {2517} (\bibinfo {year}
  {1994})}\BibitemShut {NoStop}%
\bibitem [{\citenamefont {Beck}\ and\ \citenamefont {Cohen}(2003)}]{Beck2003}%
  \BibitemOpen
  \bibfield  {author} {\bibinfo {author} {\bibfnamefont {C.}~\bibnamefont
  {Beck}}\ and\ \bibinfo {author} {\bibfnamefont {E.}~\bibnamefont {Cohen}},\
  }\href {https://doi.org/https://doi.org/10.1016/S0378-4371(03)00019-0}
  {\bibfield  {journal} {\bibinfo  {journal} {Phys. A: Stat. Mech. Appl.}\
  }\textbf {\bibinfo {volume} {322}},\ \bibinfo {pages} {267} (\bibinfo {year}
  {2003})}\BibitemShut {NoStop}%
\bibitem [{\citenamefont {Beck}\ \emph {et~al.}(2005)\citenamefont {Beck},
  \citenamefont {Cohen},\ and\ \citenamefont {Swinney}}]{Beck2005}%
  \BibitemOpen
  \bibfield  {author} {\bibinfo {author} {\bibfnamefont {C.}~\bibnamefont
  {Beck}}, \bibinfo {author} {\bibfnamefont {E.~G.~D.}\ \bibnamefont {Cohen}},\
  and\ \bibinfo {author} {\bibfnamefont {H.~L.}\ \bibnamefont {Swinney}},\
  }\href {https://doi.org/10.1103/PhysRevE.72.056133} {\bibfield  {journal}
  {\bibinfo  {journal} {Phys. Rev. E}\ }\textbf {\bibinfo {volume} {72}},\
  \bibinfo {pages} {056133} (\bibinfo {year} {2005})}\BibitemShut {NoStop}%
\bibitem [{\citenamefont {Chubynsky}\ and\ \citenamefont
  {Slater}(2014)}]{Chubynsky2014}%
  \BibitemOpen
  \bibfield  {author} {\bibinfo {author} {\bibfnamefont {M.~V.}\ \bibnamefont
  {Chubynsky}}\ and\ \bibinfo {author} {\bibfnamefont {G.~W.}\ \bibnamefont
  {Slater}},\ }\href {https://doi.org/10.1103/PhysRevLett.113.098302}
  {\bibfield  {journal} {\bibinfo  {journal} {Phys. Rev. Lett.}\ }\textbf
  {\bibinfo {volume} {113}},\ \bibinfo {pages} {098302} (\bibinfo {year}
  {2014})}\BibitemShut {NoStop}%
\bibitem [{\citenamefont {Jain}\ and\ \citenamefont
  {Sebastian}(2016)}]{doi:10.1021/acs.jpcb.6b01527}%
  \BibitemOpen
  \bibfield  {author} {\bibinfo {author} {\bibfnamefont {R.}~\bibnamefont
  {Jain}}\ and\ \bibinfo {author} {\bibfnamefont {K.~L.}\ \bibnamefont
  {Sebastian}},\ }\href {https://doi.org/10.1021/acs.jpcb.6b01527} {\bibfield
  {journal} {\bibinfo  {journal} {J. Phys. Chem. B}\ }\textbf {\bibinfo
  {volume} {120}},\ \bibinfo {pages} {3988} (\bibinfo {year}
  {2016})}\BibitemShut {NoStop}%
\bibitem [{\citenamefont {Jain}\ and\ \citenamefont
  {Sebastian}(2018)}]{PhysRevE.98.052138}%
  \BibitemOpen
  \bibfield  {author} {\bibinfo {author} {\bibfnamefont {R.}~\bibnamefont
  {Jain}}\ and\ \bibinfo {author} {\bibfnamefont {K.~L.}\ \bibnamefont
  {Sebastian}},\ }\href {https://doi.org/10.1103/PhysRevE.98.052138} {\bibfield
   {journal} {\bibinfo  {journal} {Phys. Rev. E}\ }\textbf {\bibinfo {volume}
  {98}},\ \bibinfo {pages} {052138} (\bibinfo {year} {2018})}\BibitemShut
  {NoStop}%
\bibitem [{\citenamefont {Chechkin}\ \emph {et~al.}(2017)\citenamefont
  {Chechkin}, \citenamefont {Seno}, \citenamefont {Metzler},\ and\
  \citenamefont {Sokolov}}]{PhysRevX.7.021002}%
  \BibitemOpen
  \bibfield  {author} {\bibinfo {author} {\bibfnamefont {A.~V.}\ \bibnamefont
  {Chechkin}}, \bibinfo {author} {\bibfnamefont {F.}~\bibnamefont {Seno}},
  \bibinfo {author} {\bibfnamefont {R.}~\bibnamefont {Metzler}},\ and\ \bibinfo
  {author} {\bibfnamefont {I.~M.}\ \bibnamefont {Sokolov}},\ }\href
  {https://doi.org/10.1103/PhysRevX.7.021002} {\bibfield  {journal} {\bibinfo
  {journal} {Phys. Rev. X}\ }\textbf {\bibinfo {volume} {7}},\ \bibinfo {pages}
  {021002} (\bibinfo {year} {2017})}\BibitemShut {NoStop}%
\bibitem [{\citenamefont {Lanoisel\'ee}\ \emph {et~al.}(2018)\citenamefont
  {Lanoisel\'ee}, \citenamefont {Moutal},\ and\ \citenamefont
  {Grebenkov}}]{Lanoiselee2018_FPT}%
  \BibitemOpen
  \bibfield  {author} {\bibinfo {author} {\bibfnamefont {Y.}~\bibnamefont
  {Lanoisel\'ee}}, \bibinfo {author} {\bibfnamefont {N.}~\bibnamefont
  {Moutal}},\ and\ \bibinfo {author} {\bibfnamefont {D.}~\bibnamefont
  {Grebenkov}},\ }\href
  {https://doi.org/https://doi.org/10.1038/s41467-018-06610-6} {\bibfield
  {journal} {\bibinfo  {journal} {Nat. Commun.}\ }\textbf {\bibinfo {volume}
  {9}},\ \bibinfo {pages} {1} (\bibinfo {year} {2018})}\BibitemShut {NoStop}%
\bibitem [{\citenamefont {Sposini}\ \emph {et~al.}(2018)\citenamefont
  {Sposini}, \citenamefont {Chechkin}, \citenamefont {Seno}, \citenamefont
  {Pagnini},\ and\ \citenamefont {Metzler}}]{sposini_etal-njp-2018}%
  \BibitemOpen
  \bibfield  {author} {\bibinfo {author} {\bibfnamefont {V.}~\bibnamefont
  {Sposini}}, \bibinfo {author} {\bibfnamefont {A.~V.}\ \bibnamefont
  {Chechkin}}, \bibinfo {author} {\bibfnamefont {F.}~\bibnamefont {Seno}},
  \bibinfo {author} {\bibfnamefont {G.}~\bibnamefont {Pagnini}},\ and\ \bibinfo
  {author} {\bibfnamefont {R.}~\bibnamefont {Metzler}},\ }\href
  {https://doi.org/10.1088/1367-2630/aab696} {\bibfield  {journal} {\bibinfo
  {journal} {New J. Phys.}\ }\textbf {\bibinfo {volume} {20}},\ \bibinfo
  {pages} {043044} (\bibinfo {year} {2018})}\BibitemShut {NoStop}%
\bibitem [{\citenamefont {Wang}\ \emph {et~al.}(2025)\citenamefont {Wang},
  \citenamefont {Wei}, \citenamefont {Chechkin},\ and\ \citenamefont
  {Metzler}}]{w8gv-3fxt}%
  \BibitemOpen
  \bibfield  {author} {\bibinfo {author} {\bibfnamefont {W.}~\bibnamefont
  {Wang}}, \bibinfo {author} {\bibfnamefont {Q.}~\bibnamefont {Wei}}, \bibinfo
  {author} {\bibfnamefont {A.~V.}\ \bibnamefont {Chechkin}},\ and\ \bibinfo
  {author} {\bibfnamefont {R.}~\bibnamefont {Metzler}},\ }\href
  {https://doi.org/10.1103/w8gv-3fxt} {\bibfield  {journal} {\bibinfo
  {journal} {Phys. Rev. E}\ }\textbf {\bibinfo {volume} {112}},\ \bibinfo
  {pages} {014108} (\bibinfo {year} {2025})}\BibitemShut {NoStop}%
\bibitem [{\citenamefont {Sposini}\ \emph
  {et~al.}(2024{\natexlab{a}})\citenamefont {Sposini}, \citenamefont
  {Nampoothiri}, \citenamefont {Chechkin}, \citenamefont {Orlandini},
  \citenamefont {Seno},\ and\ \citenamefont {Baldovin}}]{Sposini2024}%
  \BibitemOpen
  \bibfield  {author} {\bibinfo {author} {\bibfnamefont {V.}~\bibnamefont
  {Sposini}}, \bibinfo {author} {\bibfnamefont {S.}~\bibnamefont
  {Nampoothiri}}, \bibinfo {author} {\bibfnamefont {A.}~\bibnamefont
  {Chechkin}}, \bibinfo {author} {\bibfnamefont {E.}~\bibnamefont {Orlandini}},
  \bibinfo {author} {\bibfnamefont {F.}~\bibnamefont {Seno}},\ and\ \bibinfo
  {author} {\bibfnamefont {F.}~\bibnamefont {Baldovin}},\ }\href
  {https://doi.org/10.1103/PhysRevLett.132.117101} {\bibfield  {journal}
  {\bibinfo  {journal} {Phys. Rev. Lett.}\ }\textbf {\bibinfo {volume} {132}},\
  \bibinfo {pages} {117101} (\bibinfo {year} {2024}{\natexlab{a}})}\BibitemShut
  {NoStop}%
\bibitem [{\citenamefont {Jensen}\ \emph {et~al.}(2005)\citenamefont {Jensen},
  \citenamefont {Helpern}, \citenamefont {Ramani}, \citenamefont {Lu},\ and\
  \citenamefont {Kaczynski}}]{Jensen2005}%
  \BibitemOpen
  \bibfield  {author} {\bibinfo {author} {\bibfnamefont {J.~H.}\ \bibnamefont
  {Jensen}}, \bibinfo {author} {\bibfnamefont {J.~A.}\ \bibnamefont {Helpern}},
  \bibinfo {author} {\bibfnamefont {A.}~\bibnamefont {Ramani}}, \bibinfo
  {author} {\bibfnamefont {H.}~\bibnamefont {Lu}},\ and\ \bibinfo {author}
  {\bibfnamefont {K.}~\bibnamefont {Kaczynski}},\ }\href
  {https://doi.org/https://doi.org/10.1002/mrm.20508} {\bibfield  {journal}
  {\bibinfo  {journal} {Magn. Res. Med.}\ }\textbf {\bibinfo {volume} {53}},\
  \bibinfo {pages} {1432} (\bibinfo {year} {2005})}\BibitemShut {NoStop}%
\bibitem [{\citenamefont {Fieremans}\ \emph {et~al.}(2010)\citenamefont
  {Fieremans}, \citenamefont {Novikov}, \citenamefont {Jensen},\ and\
  \citenamefont {Helpern}}]{Fieremans2010}%
  \BibitemOpen
  \bibfield  {author} {\bibinfo {author} {\bibfnamefont {E.}~\bibnamefont
  {Fieremans}}, \bibinfo {author} {\bibfnamefont {D.~S.}\ \bibnamefont
  {Novikov}}, \bibinfo {author} {\bibfnamefont {J.~H.}\ \bibnamefont
  {Jensen}},\ and\ \bibinfo {author} {\bibfnamefont {J.~A.}\ \bibnamefont
  {Helpern}},\ }\href {https://doi.org/https://doi.org/10.1002/nbm.1577}
  {\bibfield  {journal} {\bibinfo  {journal} {NMR in Biomed.}\ }\textbf
  {\bibinfo {volume} {23}},\ \bibinfo {pages} {711} (\bibinfo {year}
  {2010})}\BibitemShut {NoStop}%
\bibitem [{\citenamefont {Grebenkov}(2019)}]{Grebenkov2019}%
  \BibitemOpen
  \bibfield  {author} {\bibinfo {author} {\bibfnamefont {D.~S.}\ \bibnamefont
  {Grebenkov}},\ }\href {https://doi.org/10.1103/PhysRevE.99.032133} {\bibfield
   {journal} {\bibinfo  {journal} {Phys. Rev. E}\ }\textbf {\bibinfo {volume}
  {99}},\ \bibinfo {pages} {032133} (\bibinfo {year} {2019})}\BibitemShut
  {NoStop}%
\bibitem [{\citenamefont {Gu\'eneau}\ \emph {et~al.}(2025)\citenamefont
  {Gu\'eneau}, \citenamefont {Majumdar},\ and\ \citenamefont
  {Schehr}}]{Gueneau2025}%
  \BibitemOpen
  \bibfield  {author} {\bibinfo {author} {\bibfnamefont {M.}~\bibnamefont
  {Gu\'eneau}}, \bibinfo {author} {\bibfnamefont {S.~N.}\ \bibnamefont
  {Majumdar}},\ and\ \bibinfo {author} {\bibfnamefont {G.}~\bibnamefont
  {Schehr}},\ }\href {https://doi.org/10.1103/rqsn-bzr6} {\bibfield  {journal}
  {\bibinfo  {journal} {Phys. Rev. Lett.}\ }\textbf {\bibinfo {volume} {135}},\
  \bibinfo {pages} {067102} (\bibinfo {year} {2025})}\BibitemShut {NoStop}%
\bibitem [{\citenamefont {Miyaguchi}\ \emph {et~al.}(2016)\citenamefont
  {Miyaguchi}, \citenamefont {Akimoto},\ and\ \citenamefont
  {Yamamoto}}]{Miyaguchi2016}%
  \BibitemOpen
  \bibfield  {author} {\bibinfo {author} {\bibfnamefont {T.}~\bibnamefont
  {Miyaguchi}}, \bibinfo {author} {\bibfnamefont {T.}~\bibnamefont {Akimoto}},\
  and\ \bibinfo {author} {\bibfnamefont {E.}~\bibnamefont {Yamamoto}},\ }\href
  {https://doi.org/10.1103/PhysRevE.94.012109} {\bibfield  {journal} {\bibinfo
  {journal} {Phys. Rev. E}\ }\textbf {\bibinfo {volume} {94}},\ \bibinfo
  {pages} {012109} (\bibinfo {year} {2016})}\BibitemShut {NoStop}%
\bibitem [{\citenamefont {Miyaguchi}\ \emph {et~al.}(2019)\citenamefont
  {Miyaguchi}, \citenamefont {Uneyama},\ and\ \citenamefont
  {Akimoto}}]{Miyaguchi2019}%
  \BibitemOpen
  \bibfield  {author} {\bibinfo {author} {\bibfnamefont {T.}~\bibnamefont
  {Miyaguchi}}, \bibinfo {author} {\bibfnamefont {T.}~\bibnamefont {Uneyama}},\
  and\ \bibinfo {author} {\bibfnamefont {T.}~\bibnamefont {Akimoto}},\ }\href
  {https://doi.org/10.1103/PhysRevE.100.012116} {\bibfield  {journal} {\bibinfo
   {journal} {Phys. Rev. E}\ }\textbf {\bibinfo {volume} {100}},\ \bibinfo
  {pages} {012116} (\bibinfo {year} {2019})}\BibitemShut {NoStop}%
\bibitem [{\citenamefont {Balcerek}\ \emph {et~al.}(2023)\citenamefont
  {Balcerek}, \citenamefont {Wyłomańska}, \citenamefont {Burnecki},
  \citenamefont {Metzler},\ and\ \citenamefont {Krapf}}]{Balcerek_2023}%
  \BibitemOpen
  \bibfield  {author} {\bibinfo {author} {\bibfnamefont {M.}~\bibnamefont
  {Balcerek}}, \bibinfo {author} {\bibfnamefont {A.}~\bibnamefont
  {Wyłomańska}}, \bibinfo {author} {\bibfnamefont {K.}~\bibnamefont
  {Burnecki}}, \bibinfo {author} {\bibfnamefont {R.}~\bibnamefont {Metzler}},\
  and\ \bibinfo {author} {\bibfnamefont {D.}~\bibnamefont {Krapf}},\ }\href
  {https://doi.org/10.1088/1367-2630/ad00d7} {\bibfield  {journal} {\bibinfo
  {journal} {New J. Phys.}\ }\textbf {\bibinfo {volume} {25}},\ \bibinfo
  {pages} {103031} (\bibinfo {year} {2023})}\BibitemShut {NoStop}%
\bibitem [{\citenamefont {Pacheco-Pozo}\ and\ \citenamefont
  {Krapf}(2024)}]{Pacheco2024}%
  \BibitemOpen
  \bibfield  {author} {\bibinfo {author} {\bibfnamefont {A.}~\bibnamefont
  {Pacheco-Pozo}}\ and\ \bibinfo {author} {\bibfnamefont {D.}~\bibnamefont
  {Krapf}},\ }\href {https://doi.org/10.1103/PhysRevE.110.014105} {\bibfield
  {journal} {\bibinfo  {journal} {Phys. Rev. E}\ }\textbf {\bibinfo {volume}
  {110}},\ \bibinfo {pages} {014105} (\bibinfo {year} {2024})}\BibitemShut
  {NoStop}%
\bibitem [{\citenamefont {Doerries}\ \emph {et~al.}(2022)\citenamefont
  {Doerries}, \citenamefont {Chechkin},\ and\ \citenamefont
  {Metzler}}]{Doerries2022}%
  \BibitemOpen
  \bibfield  {author} {\bibinfo {author} {\bibfnamefont {T.~J.}\ \bibnamefont
  {Doerries}}, \bibinfo {author} {\bibfnamefont {A.~V.}\ \bibnamefont
  {Chechkin}},\ and\ \bibinfo {author} {\bibfnamefont {R.}~\bibnamefont
  {Metzler}},\ }\href {https://doi.org/10.1098/rsif.2022.0233} {\bibfield
  {journal} {\bibinfo  {journal} {J. R. Soc. Interface}\ }\textbf {\bibinfo
  {volume} {19}},\ \bibinfo {pages} {20220233} (\bibinfo {year}
  {2022})}\BibitemShut {NoStop}%
\bibitem [{\citenamefont {Schneider}(1990)}]{schneider-1990}%
  \BibitemOpen
  \bibfield  {author} {\bibinfo {author} {\bibfnamefont {W.~R.}\ \bibnamefont
  {Schneider}},\ }in\ \href@noop {} {\emph {\bibinfo {booktitle} {Stochastic
  Processes, Physics and Geometry}}}\ (\bibinfo  {publisher} {World
  Scientific},\ \bibinfo {address} {Teaneck},\ \bibinfo {year} {1990})\ pp.\
  \bibinfo {pages} {676--681}\BibitemShut {NoStop}%
\bibitem [{\citenamefont {Schneider}(1992)}]{schneider-1992}%
  \BibitemOpen
  \bibfield  {author} {\bibinfo {author} {\bibfnamefont {W.~R.}\ \bibnamefont
  {Schneider}},\ }in\ \href@noop {} {\emph {\bibinfo {booktitle} {Ideas and
  Methods in Mathematical Analysis, Stochastics, and Applications}}},\
  Vol.~\bibinfo {volume} {I}\ (\bibinfo  {publisher} {Cambridge University
  Press},\ \bibinfo {address} {Cambridge},\ \bibinfo {year} {1992})\ pp.\
  \bibinfo {pages} {261--282}\BibitemShut {NoStop}%
\bibitem [{\citenamefont {Mura}\ and\ \citenamefont
  {Pagnini}(2008)}]{mura_etal-jpa-2008}%
  \BibitemOpen
  \bibfield  {author} {\bibinfo {author} {\bibfnamefont {A.}~\bibnamefont
  {Mura}}\ and\ \bibinfo {author} {\bibfnamefont {G.}~\bibnamefont {Pagnini}},\
  }\href {https://doi.org/10.1088/1751-8113/41/28/285003} {\bibfield  {journal}
  {\bibinfo  {journal} {J. Phys. A: Math. Theor.}\ }\textbf {\bibinfo {volume}
  {41}},\ \bibinfo {pages} {285003} (\bibinfo {year} {2008})}\BibitemShut
  {NoStop}%
\bibitem [{\citenamefont {Barkai}\ and\ \citenamefont
  {Burov}(2020)}]{Barkai2020}%
  \BibitemOpen
  \bibfield  {author} {\bibinfo {author} {\bibfnamefont {E.}~\bibnamefont
  {Barkai}}\ and\ \bibinfo {author} {\bibfnamefont {S.}~\bibnamefont {Burov}},\
  }\href {https://doi.org/10.1103/PhysRevLett.124.060603} {\bibfield  {journal}
  {\bibinfo  {journal} {Phys. Rev. Lett.}\ }\textbf {\bibinfo {volume} {124}},\
  \bibinfo {pages} {060603} (\bibinfo {year} {2020})}\BibitemShut {NoStop}%
\bibitem [{\citenamefont {Singh}\ and\ \citenamefont
  {Burov}(2024)}]{Burov2026}%
  \BibitemOpen
  \bibfield  {author} {\bibinfo {author} {\bibfnamefont {R.~K.}\ \bibnamefont
  {Singh}}\ and\ \bibinfo {author} {\bibfnamefont {S.}~\bibnamefont {Burov}},\
  }\href@noop {} {\bibinfo {title} {The emergence of laplace universality in
  correlated processes}} (\bibinfo {year} {2024}),\ \Eprint
  {https://arxiv.org/abs/2410.23112} {arXiv:2410.23112 [cond-mat.stat-mech]}
  \BibitemShut {NoStop}%
\bibitem [{\citenamefont {Akimoto}\ \emph {et~al.}(2026)\citenamefont
  {Akimoto}, \citenamefont {Jeon}, \citenamefont {Metzler}, \citenamefont
  {Miyaguchi}, \citenamefont {Uneyama},\ and\ \citenamefont
  {Yamamoto}}]{Akimoto_2026}%
  \BibitemOpen
  \bibfield  {author} {\bibinfo {author} {\bibfnamefont {T.}~\bibnamefont
  {Akimoto}}, \bibinfo {author} {\bibfnamefont {J.-H.}\ \bibnamefont {Jeon}},
  \bibinfo {author} {\bibfnamefont {R.}~\bibnamefont {Metzler}}, \bibinfo
  {author} {\bibfnamefont {T.}~\bibnamefont {Miyaguchi}}, \bibinfo {author}
  {\bibfnamefont {T.}~\bibnamefont {Uneyama}},\ and\ \bibinfo {author}
  {\bibfnamefont {E.}~\bibnamefont {Yamamoto}},\ }\href
  {https://doi.org/10.1088/1361-6633/ae358c} {\bibfield  {journal} {\bibinfo
  {journal} {Reports on Progress in Physics}\ }\textbf {\bibinfo {volume}
  {89}},\ \bibinfo {pages} {014602} (\bibinfo {year} {2026})}\BibitemShut
  {NoStop}%
\bibitem [{\citenamefont {Sliusarenko}\ \emph {et~al.}(2019)\citenamefont
  {Sliusarenko}, \citenamefont {Vitali}, \citenamefont {Sposini}, \citenamefont
  {Paradisi}, \citenamefont {Chechkin}, \citenamefont {Castellani},\ and\
  \citenamefont {Pagnini}}]{sliusarenko_etal-jpa-2019}%
  \BibitemOpen
  \bibfield  {author} {\bibinfo {author} {\bibfnamefont {O.}~\bibnamefont
  {Sliusarenko}}, \bibinfo {author} {\bibfnamefont {S.}~\bibnamefont {Vitali}},
  \bibinfo {author} {\bibfnamefont {V.}~\bibnamefont {Sposini}}, \bibinfo
  {author} {\bibfnamefont {P.}~\bibnamefont {Paradisi}}, \bibinfo {author}
  {\bibfnamefont {A.}~\bibnamefont {Chechkin}}, \bibinfo {author}
  {\bibfnamefont {G.}~\bibnamefont {Castellani}},\ and\ \bibinfo {author}
  {\bibfnamefont {G.}~\bibnamefont {Pagnini}},\ }\href
  {https://doi.org/10.1088/1751-8121/aafe90} {\bibfield  {journal} {\bibinfo
  {journal} {J. Phys. A: Math. Theor.}\ }\textbf {\bibinfo {volume} {52}},\
  \bibinfo {pages} {095601} (\bibinfo {year} {2019})}\BibitemShut {NoStop}%
\bibitem [{\citenamefont {Schulz}\ \emph {et~al.}(2013)\citenamefont {Schulz},
  \citenamefont {Barkai},\ and\ \citenamefont {Metzler}}]{Schulz2013}%
  \BibitemOpen
  \bibfield  {author} {\bibinfo {author} {\bibfnamefont {J.~H.~P.}\
  \bibnamefont {Schulz}}, \bibinfo {author} {\bibfnamefont {E.}~\bibnamefont
  {Barkai}},\ and\ \bibinfo {author} {\bibfnamefont {R.}~\bibnamefont
  {Metzler}},\ }\href {https://doi.org/10.1103/PhysRevLett.110.020602}
  {\bibfield  {journal} {\bibinfo  {journal} {Phys. Rev. Lett.}\ }\textbf
  {\bibinfo {volume} {110}},\ \bibinfo {pages} {020602} (\bibinfo {year}
  {2013})}\BibitemShut {NoStop}%
\bibitem [{\citenamefont {Barkai}(2003)}]{Barkai2003}%
  \BibitemOpen
  \bibfield  {author} {\bibinfo {author} {\bibfnamefont {E.}~\bibnamefont
  {Barkai}},\ }\href {https://doi.org/10.1103/PhysRevLett.90.104101} {\bibfield
   {journal} {\bibinfo  {journal} {Phys. Rev. Lett.}\ }\textbf {\bibinfo
  {volume} {90}},\ \bibinfo {pages} {104101} (\bibinfo {year}
  {2003})}\BibitemShut {NoStop}%
\bibitem [{\citenamefont {Molina-Garc\'ia}\ \emph {et~al.}(2016)\citenamefont
  {Molina-Garc\'ia}, \citenamefont {{Minh Pham}}, \citenamefont {Paradisi},
  \citenamefont {Manzo},\ and\ \citenamefont {Pagnini}}]{molina_etal-pre-2016}%
  \BibitemOpen
  \bibfield  {author} {\bibinfo {author} {\bibfnamefont {D.}~\bibnamefont
  {Molina-Garc\'ia}}, \bibinfo {author} {\bibfnamefont {T.}~\bibnamefont {{Minh
  Pham}}}, \bibinfo {author} {\bibfnamefont {P.}~\bibnamefont {Paradisi}},
  \bibinfo {author} {\bibfnamefont {C.}~\bibnamefont {Manzo}},\ and\ \bibinfo
  {author} {\bibfnamefont {G.}~\bibnamefont {Pagnini}},\ }\href
  {https://doi.org/https://doi.org/10.1103/PhysRevE.94.052147} {\bibfield
  {journal} {\bibinfo  {journal} {Phys. Rev. E}\ }\textbf {\bibinfo {volume}
  {94}},\ \bibinfo {pages} {052147} (\bibinfo {year} {2016})}\BibitemShut
  {NoStop}%
\bibitem [{\citenamefont {Runfola}\ and\ \citenamefont
  {Pagnini}(2024)}]{runfola_etal-pd-2024}%
  \BibitemOpen
  \bibfield  {author} {\bibinfo {author} {\bibfnamefont {C.}~\bibnamefont
  {Runfola}}\ and\ \bibinfo {author} {\bibfnamefont {G.}~\bibnamefont
  {Pagnini}},\ }\href
  {https://doi.org/https://doi.org/10.1016/j.physd.2024.134247} {\bibfield
  {journal} {\bibinfo  {journal} {Physica D}\ }\textbf {\bibinfo {volume}
  {467}},\ \bibinfo {pages} {134247} (\bibinfo {year} {2024})}\BibitemShut
  {NoStop}%
\bibitem [{\citenamefont {Massignan}\ \emph {et~al.}(2014)\citenamefont
  {Massignan}, \citenamefont {Manzo}, \citenamefont {Torreno-Pina},
  \citenamefont {Garc\'{\i}a-Parajo}, \citenamefont {Lewenstein},\ and\
  \citenamefont {Lapeyre}}]{Massignan2014}%
  \BibitemOpen
  \bibfield  {author} {\bibinfo {author} {\bibfnamefont {P.}~\bibnamefont
  {Massignan}}, \bibinfo {author} {\bibfnamefont {C.}~\bibnamefont {Manzo}},
  \bibinfo {author} {\bibfnamefont {J.~A.}\ \bibnamefont {Torreno-Pina}},
  \bibinfo {author} {\bibfnamefont {M.~F.}\ \bibnamefont {Garc\'{\i}a-Parajo}},
  \bibinfo {author} {\bibfnamefont {M.}~\bibnamefont {Lewenstein}},\ and\
  \bibinfo {author} {\bibfnamefont {G.~J.}\ \bibnamefont {Lapeyre}},\ }\href
  {https://doi.org/10.1103/PhysRevLett.112.150603} {\bibfield  {journal}
  {\bibinfo  {journal} {Phys. Rev. Lett.}\ }\textbf {\bibinfo {volume} {112}},\
  \bibinfo {pages} {150603} (\bibinfo {year} {2014})}\BibitemShut {NoStop}%
\bibitem [{\citenamefont {Manzo}\ \emph {et~al.}(2015)\citenamefont {Manzo},
  \citenamefont {Torreno-Pina}, \citenamefont {Massignan}, \citenamefont
  {Lapeyre}, \citenamefont {Lewenstein},\ and\ \citenamefont
  {Garcia~Parajo}}]{Manzo2015}%
  \BibitemOpen
  \bibfield  {author} {\bibinfo {author} {\bibfnamefont {C.}~\bibnamefont
  {Manzo}}, \bibinfo {author} {\bibfnamefont {J.~A.}\ \bibnamefont
  {Torreno-Pina}}, \bibinfo {author} {\bibfnamefont {P.}~\bibnamefont
  {Massignan}}, \bibinfo {author} {\bibfnamefont {G.~J.}\ \bibnamefont
  {Lapeyre}}, \bibinfo {author} {\bibfnamefont {M.}~\bibnamefont
  {Lewenstein}},\ and\ \bibinfo {author} {\bibfnamefont {M.~F.}\ \bibnamefont
  {Garcia~Parajo}},\ }\href {https://doi.org/10.1103/PhysRevX.5.011021}
  {\bibfield  {journal} {\bibinfo  {journal} {Phys. Rev. X}\ }\textbf {\bibinfo
  {volume} {5}},\ \bibinfo {pages} {011021} (\bibinfo {year}
  {2015})}\BibitemShut {NoStop}%
\bibitem [{\citenamefont {Akimoto}\ and\ \citenamefont
  {Yamamoto}(2016)}]{Akimoto_2016}%
  \BibitemOpen
  \bibfield  {author} {\bibinfo {author} {\bibfnamefont {T.}~\bibnamefont
  {Akimoto}}\ and\ \bibinfo {author} {\bibfnamefont {E.}~\bibnamefont
  {Yamamoto}},\ }\href {https://doi.org/10.1088/1742-5468/2016/12/123201}
  {\bibfield  {journal} {\bibinfo  {journal} {J. Stat. Mech.: Theo. Exp.}\
  }\textbf {\bibinfo {volume} {2016}},\ \bibinfo {pages} {123201} (\bibinfo
  {year} {2016})}\BibitemShut {NoStop}%
\bibitem [{\citenamefont {Kozachenko}\ and\ \citenamefont
  {Moklyachuk}(1999)}]{Kozachenko1999}%
  \BibitemOpen
  \bibfield  {author} {\bibinfo {author} {\bibfnamefont {Y.}~\bibnamefont
  {Kozachenko}}\ and\ \bibinfo {author} {\bibfnamefont {O.}~\bibnamefont
  {Moklyachuk}},\ }\href {https://doi.org/10.1023/A:1009907019950} {\bibfield
  {journal} {\bibinfo  {journal} {Extremes}\ }\textbf {\bibinfo {volume} {2}},\
  \bibinfo {pages} {269–293} (\bibinfo {year} {1999})}\BibitemShut {NoStop}%
\bibitem [{\citenamefont {Sposini}\ \emph
  {et~al.}(2024{\natexlab{b}})\citenamefont {Sposini}, \citenamefont
  {Nampoothiri}, \citenamefont {Chechkin}, \citenamefont {Orlandini},
  \citenamefont {Seno},\ and\ \citenamefont {Baldovin}}]{Sposini2024a}%
  \BibitemOpen
  \bibfield  {author} {\bibinfo {author} {\bibfnamefont {V.}~\bibnamefont
  {Sposini}}, \bibinfo {author} {\bibfnamefont {S.}~\bibnamefont
  {Nampoothiri}}, \bibinfo {author} {\bibfnamefont {A.}~\bibnamefont
  {Chechkin}}, \bibinfo {author} {\bibfnamefont {E.}~\bibnamefont {Orlandini}},
  \bibinfo {author} {\bibfnamefont {F.}~\bibnamefont {Seno}},\ and\ \bibinfo
  {author} {\bibfnamefont {F.}~\bibnamefont {Baldovin}},\ }\href
  {https://doi.org/10.1103/PhysRevLett.132.117101} {\bibfield  {journal}
  {\bibinfo  {journal} {Phys. Rev. Lett.}\ }\textbf {\bibinfo {volume} {132}},\
  \bibinfo {pages} {117101} (\bibinfo {year} {2024}{\natexlab{b}})}\BibitemShut
  {NoStop}%
\bibitem [{\citenamefont {Sposini}\ \emph
  {et~al.}(2024{\natexlab{c}})\citenamefont {Sposini}, \citenamefont
  {Nampoothiri}, \citenamefont {Chechkin}, \citenamefont {Orlandini},
  \citenamefont {Seno},\ and\ \citenamefont {Baldovin}}]{Sposini2024b}%
  \BibitemOpen
  \bibfield  {author} {\bibinfo {author} {\bibfnamefont {V.}~\bibnamefont
  {Sposini}}, \bibinfo {author} {\bibfnamefont {S.}~\bibnamefont
  {Nampoothiri}}, \bibinfo {author} {\bibfnamefont {A.}~\bibnamefont
  {Chechkin}}, \bibinfo {author} {\bibfnamefont {E.}~\bibnamefont {Orlandini}},
  \bibinfo {author} {\bibfnamefont {F.}~\bibnamefont {Seno}},\ and\ \bibinfo
  {author} {\bibfnamefont {F.}~\bibnamefont {Baldovin}},\ }\href
  {https://doi.org/10.1103/PhysRevE.109.034120} {\bibfield  {journal} {\bibinfo
   {journal} {Phys. Rev. E}\ }\textbf {\bibinfo {volume} {109}},\ \bibinfo
  {pages} {034120} (\bibinfo {year} {2024}{\natexlab{c}})}\BibitemShut
  {NoStop}%
\bibitem [{\citenamefont {Lanoisel\'ee}\ and\ \citenamefont
  {Grebenkov}(2018)}]{Lanoiselee2018_NGmodel}%
  \BibitemOpen
  \bibfield  {author} {\bibinfo {author} {\bibfnamefont {Y.}~\bibnamefont
  {Lanoisel\'ee}}\ and\ \bibinfo {author} {\bibfnamefont {D.}~\bibnamefont
  {Grebenkov}},\ }\href {https://doi.org/10.1088/1751-8121/aab15f} {\bibfield
  {journal} {\bibinfo  {journal} {J. Phys. A: Math. Theo.}\ }\textbf {\bibinfo
  {volume} {51}},\ \bibinfo {pages} {145602} (\bibinfo {year}
  {2018})}\BibitemShut {NoStop}%
\bibitem [{\citenamefont {Magdziarz}\ \emph {et~al.}(2008)\citenamefont
  {Magdziarz}, \citenamefont {Weron},\ and\ \citenamefont
  {Klafter}}]{Magdziarz2008}%
  \BibitemOpen
  \bibfield  {author} {\bibinfo {author} {\bibfnamefont {M.}~\bibnamefont
  {Magdziarz}}, \bibinfo {author} {\bibfnamefont {A.}~\bibnamefont {Weron}},\
  and\ \bibinfo {author} {\bibfnamefont {J.}~\bibnamefont {Klafter}},\ }\href
  {https://doi.org/10.1103/PhysRevLett.101.210601} {\bibfield  {journal}
  {\bibinfo  {journal} {Phys. Rev. Lett.}\ }\textbf {\bibinfo {volume} {101}},\
  \bibinfo {pages} {210601} (\bibinfo {year} {2008})}\BibitemShut {NoStop}%
\bibitem [{\citenamefont {Vitali}\ \emph {et~al.}(2018)\citenamefont {Vitali},
  \citenamefont {Sposini}, \citenamefont {Sliusarenko}, \citenamefont
  {Paradisi}, \citenamefont {Castellani},\ and\ \citenamefont
  {Pagnini}}]{vitali_etal-jrsi-2018}%
  \BibitemOpen
  \bibfield  {author} {\bibinfo {author} {\bibfnamefont {S.}~\bibnamefont
  {Vitali}}, \bibinfo {author} {\bibfnamefont {V.}~\bibnamefont {Sposini}},
  \bibinfo {author} {\bibfnamefont {O.}~\bibnamefont {Sliusarenko}}, \bibinfo
  {author} {\bibfnamefont {P.}~\bibnamefont {Paradisi}}, \bibinfo {author}
  {\bibfnamefont {G.}~\bibnamefont {Castellani}},\ and\ \bibinfo {author}
  {\bibfnamefont {G.}~\bibnamefont {Pagnini}},\ }\href
  {https://doi.org/https://doi.org/10.1098/rsif.2018.0282} {\bibfield
  {journal} {\bibinfo  {journal} {J. R. Soc. Interface}\ }\textbf {\bibinfo
  {volume} {15}},\ \bibinfo {pages} {20180282} (\bibinfo {year}
  {2018})}\BibitemShut {NoStop}%
\bibitem [{\citenamefont {Lanoisel\'ee}\ \emph {et~al.}(2022)\citenamefont
  {Lanoisel\'ee}, \citenamefont {Stanislavsky}, \citenamefont {Calebiro},\ and\
  \citenamefont {Weron}}]{Lanoiselee2023}%
  \BibitemOpen
  \bibfield  {author} {\bibinfo {author} {\bibfnamefont {Y.}~\bibnamefont
  {Lanoisel\'ee}}, \bibinfo {author} {\bibfnamefont {A.}~\bibnamefont
  {Stanislavsky}}, \bibinfo {author} {\bibfnamefont {D.}~\bibnamefont
  {Calebiro}},\ and\ \bibinfo {author} {\bibfnamefont {A.}~\bibnamefont
  {Weron}},\ }\href {https://doi.org/10.1103/PhysRevE.106.064127} {\bibfield
  {journal} {\bibinfo  {journal} {Phys. Rev. E}\ }\textbf {\bibinfo {volume}
  {106}},\ \bibinfo {pages} {064127} (\bibinfo {year} {2022})}\BibitemShut
  {NoStop}%
\bibitem [{\citenamefont {Wang}\ \emph
  {et~al.}(2020{\natexlab{a}})\citenamefont {Wang}, \citenamefont {Cherstvy},
  \citenamefont {Chechkin}, \citenamefont {Thapa}, \citenamefont {Seno},
  \citenamefont {Liu},\ and\ \citenamefont {Metzler}}]{Wang2020}%
  \BibitemOpen
  \bibfield  {author} {\bibinfo {author} {\bibfnamefont {W.}~\bibnamefont
  {Wang}}, \bibinfo {author} {\bibfnamefont {A.~G.}\ \bibnamefont {Cherstvy}},
  \bibinfo {author} {\bibfnamefont {A.~V.}\ \bibnamefont {Chechkin}}, \bibinfo
  {author} {\bibfnamefont {S.}~\bibnamefont {Thapa}}, \bibinfo {author}
  {\bibfnamefont {F.}~\bibnamefont {Seno}}, \bibinfo {author} {\bibfnamefont
  {X.}~\bibnamefont {Liu}},\ and\ \bibinfo {author} {\bibfnamefont
  {R.}~\bibnamefont {Metzler}},\ }\href
  {https://doi.org/10.1088/1751-8121/aba467} {\bibfield  {journal} {\bibinfo
  {journal} {J. Phys. A: Math. Theo.}\ }\textbf {\bibinfo {volume} {53}},\
  \bibinfo {pages} {474001} (\bibinfo {year} {2020}{\natexlab{a}})}\BibitemShut
  {NoStop}%
\bibitem [{\citenamefont {Burnecki}\ and\ \citenamefont
  {Weron}(2010)}]{Burnecki2010}%
  \BibitemOpen
  \bibfield  {author} {\bibinfo {author} {\bibfnamefont {K.}~\bibnamefont
  {Burnecki}}\ and\ \bibinfo {author} {\bibfnamefont {A.}~\bibnamefont
  {Weron}},\ }\href {https://doi.org/10.1103/PhysRevE.82.021130} {\bibfield
  {journal} {\bibinfo  {journal} {Phys. Rev. E}\ }\textbf {\bibinfo {volume}
  {82}},\ \bibinfo {pages} {021130} (\bibinfo {year} {2010})}\BibitemShut
  {NoStop}%
\bibitem [{\citenamefont {Maller}\ \emph {et~al.}(2009)\citenamefont {Maller},
  \citenamefont {M{\"u}ller},\ and\ \citenamefont {Szimayer}}]{Maller2009}%
  \BibitemOpen
  \bibfield  {author} {\bibinfo {author} {\bibfnamefont {R.~A.}\ \bibnamefont
  {Maller}}, \bibinfo {author} {\bibfnamefont {G.}~\bibnamefont {M{\"u}ller}},\
  and\ \bibinfo {author} {\bibfnamefont {A.}~\bibnamefont {Szimayer}},\
  }\bibinfo {title} {Ornstein--uhlenbeck processes and extensions},\ in\ \href
  {https://doi.org/10.1007/978-3-540-71297-8_18} {\emph {\bibinfo {booktitle}
  {Handbook of Financial Time Series}}},\ \bibinfo {editor} {edited by\
  \bibinfo {editor} {\bibfnamefont {T.}~\bibnamefont {Mikosch}}, \bibinfo
  {editor} {\bibfnamefont {J.-P.}\ \bibnamefont {Krei{\ss}}}, \bibinfo {editor}
  {\bibfnamefont {R.~A.}\ \bibnamefont {Davis}},\ and\ \bibinfo {editor}
  {\bibfnamefont {T.~G.}\ \bibnamefont {Andersen}}}\ (\bibinfo  {publisher}
  {Springer Berlin Heidelberg},\ \bibinfo {address} {Berlin, Heidelberg},\
  \bibinfo {year} {2009})\ pp.\ \bibinfo {pages} {421--437}\BibitemShut
  {NoStop}%
\bibitem [{\citenamefont {Wang}\ \emph
  {et~al.}(2020{\natexlab{b}})\citenamefont {Wang}, \citenamefont {Barkai},\
  and\ \citenamefont {Burov}}]{Wang2020b}%
  \BibitemOpen
  \bibfield  {author} {\bibinfo {author} {\bibfnamefont {W.}~\bibnamefont
  {Wang}}, \bibinfo {author} {\bibfnamefont {E.}~\bibnamefont {Barkai}},\ and\
  \bibinfo {author} {\bibfnamefont {S.}~\bibnamefont {Burov}},\ }\href@noop {}
  {\bibfield  {journal} {\bibinfo  {journal} {Entropy}\ }\textbf {\bibinfo
  {volume} {22}} (\bibinfo {year} {2020}{\natexlab{b}})}\BibitemShut {NoStop}%
\bibitem [{\citenamefont {Lanoisel\'ee}\ and\ \citenamefont
  {Grebenkov}(2019)}]{Lanoiselee2019}%
  \BibitemOpen
  \bibfield  {author} {\bibinfo {author} {\bibfnamefont {Y.}~\bibnamefont
  {Lanoisel\'ee}}\ and\ \bibinfo {author} {\bibfnamefont {D.}~\bibnamefont
  {Grebenkov}},\ }\href {https://doi.org/10.1088/1751-8121/ab2826} {\bibfield
  {journal} {\bibinfo  {journal} {J. Phys. A: Math. Theo.}\ }\textbf {\bibinfo
  {volume} {52}},\ \bibinfo {pages} {304001} (\bibinfo {year}
  {2019})}\BibitemShut {NoStop}%
\bibitem [{\citenamefont {Alexandre}\ \emph {et~al.}(2023)\citenamefont
  {Alexandre}, \citenamefont {Lavaud}, \citenamefont {Fares}, \citenamefont
  {Millan}, \citenamefont {Louyer}, \citenamefont {Salez}, \citenamefont
  {Amarouchene}, \citenamefont {Gu\'erin},\ and\ \citenamefont
  {Dean}}]{Alexandre2023}%
  \BibitemOpen
  \bibfield  {author} {\bibinfo {author} {\bibfnamefont {A.}~\bibnamefont
  {Alexandre}}, \bibinfo {author} {\bibfnamefont {M.}~\bibnamefont {Lavaud}},
  \bibinfo {author} {\bibfnamefont {N.}~\bibnamefont {Fares}}, \bibinfo
  {author} {\bibfnamefont {E.}~\bibnamefont {Millan}}, \bibinfo {author}
  {\bibfnamefont {Y.}~\bibnamefont {Louyer}}, \bibinfo {author} {\bibfnamefont
  {T.}~\bibnamefont {Salez}}, \bibinfo {author} {\bibfnamefont
  {Y.}~\bibnamefont {Amarouchene}}, \bibinfo {author} {\bibfnamefont
  {T.}~\bibnamefont {Gu\'erin}},\ and\ \bibinfo {author} {\bibfnamefont
  {D.~S.}\ \bibnamefont {Dean}},\ }\href
  {https://doi.org/10.1103/PhysRevLett.130.077101} {\bibfield  {journal}
  {\bibinfo  {journal} {Phys. Rev. Lett.}\ }\textbf {\bibinfo {volume} {130}},\
  \bibinfo {pages} {077101} (\bibinfo {year} {2023})}\BibitemShut {NoStop}%
\bibitem [{\citenamefont {Grebenkov}(2011)}]{Grebenkov2011b}%
  \BibitemOpen
  \bibfield  {author} {\bibinfo {author} {\bibfnamefont {D.~S.}\ \bibnamefont
  {Grebenkov}},\ }\href {https://doi.org/10.1103/PhysRevE.84.031124} {\bibfield
   {journal} {\bibinfo  {journal} {Phys. Rev. E}\ }\textbf {\bibinfo {volume}
  {84}},\ \bibinfo {pages} {031124} (\bibinfo {year} {2011})}\BibitemShut
  {NoStop}%
\bibitem [{\citenamefont {Lanoisel\'ee}\ \emph {et~al.}(2025)\citenamefont
  {Lanoisel\'ee}, \citenamefont {Pagnini},\ and\ \citenamefont
  {Wy\l{}oma\ifmmode~\acute{n}\else \'{n}\fi{}ska}}]{Lanoiselee2025}%
  \BibitemOpen
  \bibfield  {author} {\bibinfo {author} {\bibfnamefont {Y.}~\bibnamefont
  {Lanoisel\'ee}}, \bibinfo {author} {\bibfnamefont {G.}~\bibnamefont
  {Pagnini}},\ and\ \bibinfo {author} {\bibfnamefont {A.}~\bibnamefont
  {Wy\l{}oma\ifmmode~\acute{n}\else \'{n}\fi{}ska}},\ }\href
  {https://doi.org/10.1103/y5pn-5ynd} {\bibfield  {journal} {\bibinfo
  {journal} {Phys. Rev. Lett.}\ }\textbf {\bibinfo {volume} {135}},\ \bibinfo
  {pages} {137101} (\bibinfo {year} {2025})}\BibitemShut {NoStop}%
\bibitem [{\citenamefont {Wyłomańska}\ \emph {et~al.}(2015)\citenamefont
  {Wyłomańska}, \citenamefont {Chechkin}, \citenamefont {Gajda},\ and\
  \citenamefont {Sokolov}}]{WYLOMANSKA2015}%
  \BibitemOpen
  \bibfield  {author} {\bibinfo {author} {\bibfnamefont {A.}~\bibnamefont
  {Wyłomańska}}, \bibinfo {author} {\bibfnamefont {A.}~\bibnamefont
  {Chechkin}}, \bibinfo {author} {\bibfnamefont {J.}~\bibnamefont {Gajda}},\
  and\ \bibinfo {author} {\bibfnamefont {I.~M.}\ \bibnamefont {Sokolov}},\
  }\href {https://doi.org/https://doi.org/10.1016/j.physa.2014.11.049}
  {\bibfield  {journal} {\bibinfo  {journal} {Physica A: Statistical Mechanics
  and its Applications}\ }\textbf {\bibinfo {volume} {421}},\ \bibinfo {pages}
  {412} (\bibinfo {year} {2015})}\BibitemShut {NoStop}%
\bibitem [{\citenamefont {Pacheco-Pozo}\ and\ \citenamefont
  {Sokolov}(2023)}]{Pacheco2023}%
  \BibitemOpen
  \bibfield  {author} {\bibinfo {author} {\bibfnamefont {A.}~\bibnamefont
  {Pacheco-Pozo}}\ and\ \bibinfo {author} {\bibfnamefont {I.~M.}\ \bibnamefont
  {Sokolov}},\ }\href
  {https://doi.org/https://doi.org/10.1140/epjb/s10051-023-00621-z} {\bibfield
  {journal} {\bibinfo  {journal} {Eur. Phys. J. B}\ }\textbf {\bibinfo {volume}
  {96}},\ \bibinfo {pages} {152} (\bibinfo {year} {2023})}\BibitemShut
  {NoStop}%
\bibitem [{\citenamefont {Bouchaud}\ and\ \citenamefont
  {Georges}(1990)}]{Bouchaud1990}%
  \BibitemOpen
  \bibfield  {author} {\bibinfo {author} {\bibfnamefont {J.-P.}\ \bibnamefont
  {Bouchaud}}\ and\ \bibinfo {author} {\bibfnamefont {A.}~\bibnamefont
  {Georges}},\ }\href
  {https://doi.org/https://doi.org/10.1016/0370-1573(90)90099-N} {\bibfield
  {journal} {\bibinfo  {journal} {Physics Reports}\ }\textbf {\bibinfo {volume}
  {195}},\ \bibinfo {pages} {127} (\bibinfo {year} {1990})}\BibitemShut
  {NoStop}%
\bibitem [{\citenamefont {Pagnini}\ and\ \citenamefont
  {Paradisi}(2016)}]{pagnini_etal-fcaa-2016}%
  \BibitemOpen
  \bibfield  {author} {\bibinfo {author} {\bibfnamefont {G.}~\bibnamefont
  {Pagnini}}\ and\ \bibinfo {author} {\bibfnamefont {P.}~\bibnamefont
  {Paradisi}},\ }\href {https://doi.org/https://doi.org/10.1515/fca-2016-0022}
  {\bibfield  {journal} {\bibinfo  {journal} {Fract. Calc. Appl. Anal.}\
  }\textbf {\bibinfo {volume} {19}},\ \bibinfo {pages} {408} (\bibinfo {year}
  {2016})}\BibitemShut {NoStop}%
\bibitem [{\citenamefont {Pacheco-Pozo}\ and\ \citenamefont
  {Sokolov}(2021)}]{Pacheco2024b}%
  \BibitemOpen
  \bibfield  {author} {\bibinfo {author} {\bibfnamefont {A.}~\bibnamefont
  {Pacheco-Pozo}}\ and\ \bibinfo {author} {\bibfnamefont {I.~M.}\ \bibnamefont
  {Sokolov}},\ }\href {https://doi.org/10.1103/PhysRevE.103.042116} {\bibfield
  {journal} {\bibinfo  {journal} {Phys. Rev. E}\ }\textbf {\bibinfo {volume}
  {103}},\ \bibinfo {pages} {042116} (\bibinfo {year} {2021})}\BibitemShut
  {NoStop}%
\bibitem [{\citenamefont {Laloux}\ \emph {et~al.}(2000)\citenamefont {Laloux},
  \citenamefont {Cizeau}, \citenamefont {Potters},\ and\ \citenamefont
  {Bouchaud}}]{Laloux2000}%
  \BibitemOpen
  \bibfield  {author} {\bibinfo {author} {\bibfnamefont {L.}~\bibnamefont
  {Laloux}}, \bibinfo {author} {\bibfnamefont {P.}~\bibnamefont {Cizeau}},
  \bibinfo {author} {\bibfnamefont {M.}~\bibnamefont {Potters}},\ and\ \bibinfo
  {author} {\bibfnamefont {J.-P.}\ \bibnamefont {Bouchaud}},\ }\href
  {https://doi.org/10.1142/S0219024900000255} {\bibfield  {journal} {\bibinfo
  {journal} {Int. J. Theo. Appl. Fin.}\ }\textbf {\bibinfo {volume} {03}},\
  \bibinfo {pages} {391} (\bibinfo {year} {2000})}\BibitemShut {NoStop}%
\bibitem [{\citenamefont {Lamrani}\ \emph {et~al.}(2025)\citenamefont
  {Lamrani}, \citenamefont {Collins},\ and\ \citenamefont
  {Bouchaud}}]{Lamrani2025}%
  \BibitemOpen
  \bibfield  {author} {\bibinfo {author} {\bibfnamefont {L.}~\bibnamefont
  {Lamrani}}, \bibinfo {author} {\bibfnamefont {B.}~\bibnamefont {Collins}},\
  and\ \bibinfo {author} {\bibfnamefont {J.-P.}\ \bibnamefont {Bouchaud}},\
  }\href {https://doi.org/10.48550/arXiv.2509.13923} {\bibinfo {title} {Holdout
  cross-validation for large non-gaussian covariance matrix estimation using
  weingarten calculus}} (\bibinfo {year} {2025}),\ \Eprint
  {https://arxiv.org/abs/2509.13923} {arXiv:2509.13923 [q-fin.ST]} \BibitemShut
  {NoStop}%
\bibitem [{\citenamefont {Balcerek}\ \emph {et~al.}(2025)\citenamefont
  {Balcerek}, \citenamefont {Thapa}, \citenamefont {Burnecki}, \citenamefont
  {Kantz}, \citenamefont {Metzler}, \citenamefont
  {Wy\l{}oma\ifmmode~\acute{n}\else \'{n}\fi{}ska},\ and\ \citenamefont
  {Chechkin}}]{Balcerek2025}%
  \BibitemOpen
  \bibfield  {author} {\bibinfo {author} {\bibfnamefont {M.}~\bibnamefont
  {Balcerek}}, \bibinfo {author} {\bibfnamefont {S.}~\bibnamefont {Thapa}},
  \bibinfo {author} {\bibfnamefont {K.}~\bibnamefont {Burnecki}}, \bibinfo
  {author} {\bibfnamefont {H.}~\bibnamefont {Kantz}}, \bibinfo {author}
  {\bibfnamefont {R.}~\bibnamefont {Metzler}}, \bibinfo {author} {\bibfnamefont
  {A.}~\bibnamefont {Wy\l{}oma\ifmmode~\acute{n}\else \'{n}\fi{}ska}},\ and\
  \bibinfo {author} {\bibfnamefont {A.}~\bibnamefont {Chechkin}},\ }\href
  {https://doi.org/10.1103/PhysRevLett.134.197101} {\bibfield  {journal}
  {\bibinfo  {journal} {Phys. Rev. Lett.}\ }\textbf {\bibinfo {volume} {134}},\
  \bibinfo {pages} {197101} (\bibinfo {year} {2025})}\BibitemShut {NoStop}%
\bibitem [{\citenamefont {Lanoisel\'ee}\ and\ \citenamefont
  {Grebenkov}(2016)}]{Lanoiselee2016}%
  \BibitemOpen
  \bibfield  {author} {\bibinfo {author} {\bibfnamefont {Y.}~\bibnamefont
  {Lanoisel\'ee}}\ and\ \bibinfo {author} {\bibfnamefont {D.~S.}\ \bibnamefont
  {Grebenkov}},\ }\href {https://doi.org/10.1103/PhysRevE.93.052146} {\bibfield
   {journal} {\bibinfo  {journal} {Phys. Rev. E}\ }\textbf {\bibinfo {volume}
  {93}},\ \bibinfo {pages} {052146} (\bibinfo {year} {2016})}\BibitemShut
  {NoStop}%
\bibitem [{\citenamefont {Cox}\ \emph {et~al.}(1985)\citenamefont {Cox},
  \citenamefont {Ingersoll},\ and\ \citenamefont {Ross}}]{Cox1985}%
  \BibitemOpen
  \bibfield  {author} {\bibinfo {author} {\bibfnamefont {J.~C.}\ \bibnamefont
  {Cox}}, \bibinfo {author} {\bibfnamefont {J.~E.}\ \bibnamefont {Ingersoll}},\
  and\ \bibinfo {author} {\bibfnamefont {S.~A.}\ \bibnamefont {Ross}},\ }\href
  {https://doi.org/https://doi.org/10.2307/1911242} {\bibfield  {journal}
  {\bibinfo  {journal} {Econometrica}\ }\textbf {\bibinfo {volume} {53}},\
  \bibinfo {pages} {385} (\bibinfo {year} {1985})}\BibitemShut {NoStop}%
\end{thebibliography}

\end{document}